\documentclass[onefignum,onetabnum]{siamart190516}

\usepackage{lipsum}
\usepackage{amsfonts}
\usepackage{graphicx}
\usepackage{epstopdf}
\usepackage{algorithmic}
\ifpdf
  \DeclareGraphicsExtensions{.eps,.pdf,.png,.jpg}
\else
  \DeclareGraphicsExtensions{.eps}
\fi

\usepackage{enumitem}
\setlist[enumerate]{leftmargin=.5in}
\setlist[itemize]{leftmargin=.5in}

\newsiamremark{remark}{Remark}
\newsiamremark{hypothesis}{Hypothesis}
\crefname{hypothesis}{Hypothesis}{Hypotheses}
\newsiamthm{claim}{Claim}

\headers{A Mathematical Model for the Origin of Name Brands and Generics}{J. D. Johnson, A. M. Redlich, and D. M. Abrams}

\usepackage{overpic}
\usepackage{subcaption}
\usepackage{pgfplots}
\usepackage{graphicx}
\usepackage{tikz}
\usepackage{environ}
\usepackage[normalem]{ulem} 
\usepackage{mathptmx}
\usepackage{enumitem}

\makeatletter
\newsavebox{\measure@tikzpicture}
\NewEnviron{scaletikzpicturetowidth}[1]{%
  \def\tikz@width{#1}%
  \begin{lrbox}{\measure@tikzpicture}%
  \BODY
  \end{lrbox}%
  \pgfmathparse{#1/\wd\measure@tikzpicture}%
  \BODY
}
\makeatother

\usetikzlibrary{calc,angles,positioning,intersections}
\usetikzlibrary{decorations.pathreplacing,quotes}
\usetikzlibrary{calc,angles,positioning,intersections}

\usepgfplotslibrary{fillbetween}
\usepgfplotslibrary{groupplots}
\pgfplotsset{compat = newest}

\graphicspath{{figs/}}

\DeclareMathOperator{\rma}{\mathrm{a}}
\DeclareMathOperator{\rmQ}{\mathrm{Q}}

\newcommand{\xcrit}{x_{\text{crit}}}

\newcommand{\abar}{\overline{a}}
\newcommand{\Qfree}{Q_{\mathrm{free}}(a_i|\vec{a})}
\newcommand{\kP}{k_{\mathrm{P}}}
\newcommand{\kQ}{k_{\mathrm{Q}}}
\newcommand{\kAD}{\Delta Q_\textrm{ad}}
\newcommand{\ka}{k_{\mathrm{a}}}
\newcommand{\Qmin}{Q_{\mathrm{min}}}
\newcommand{\FigBeg}{Figure~}
\newcommand{\FigMid}{Fig.~}

\newcommand{\EqMid}{Eq.~}
\newcommand{\Qstar}{Q_i^*}
\newcommand{\Pstar}{P_i^*}
\newcommand{\width}{\lambda}
\newcommand{\MB}{\mathrm{B}}

\setlist{topsep=8pt,itemsep=4pt,partopsep=4pt, parsep=4pt}

\title{A Mathematical Model for the Origin of Name Brands and Generics\thanks{Submitted to the editors DATE.
\funding{This work was funded by the National Science Foundation support through Research Training Grant No. 1547394}}}

\author{Joseph D. Johnson\thanks{Department of Engineering Sciences and Applied Mathematics, Northwestern University (josephjohnson2020@u.northwestern.edu)}
\and Adam M. Redlich\footnotemark[2]
\and Daniel M. Abrams\thanks{Department of Engineering Sciences and Applied Mathematics, and Northwestern Institute on
Complex Systems, Northwestern University, Evanston, IL 60208 (dmabrams@northwestern.edu).
}}

\usepackage{amsopn}

\ifpdf
\hypersetup{
  pdftitle={An Example Article},
  pdfauthor={D. Doe, P. T. Frank, and J. E. Smith}
}
\fi

\begin{document}

\maketitle

\begin{abstract}
Firms in the U.S.~spend over 200 billion dollars each year advertising their products to consumers, around one percent of the country's gross domestic product. It is of great interest to understand how that aggregate expenditure affects prices, market efficiency, and overall welfare. 
Here, we present a mathematical model for the dynamics of competition through advertising and find a surprising prediction: when advertising is relatively cheap compared to the maximum benefit advertising offers, rational firms split into two groups, one with significantly less advertising (a ``generic'' group) and one with significantly more advertising (a ``name brand'' group). 
Our model predicts that this segmentation will also be reflected in price distributions; we use large consumer data sets to test this prediction and find good qualitative agreement. 
\end{abstract}




	










\section{Introduction and background}

Advertising is an important component of a free market system; it has been estimated that advertising expenditures in the United States exceeded \$200 billion dollars in 2018 alone \cite{shaban_2019}. Although the monetary investment is large, it remains unclear exactly how advertising affects demand and what the implications are for market competition. Perhaps advertising leads to increased market efficiency, greater aggregate profit for sellers, or better outcomes for buyers. The opposite could also be argued.

There are three prevailing theories as to how advertising influences the consumer \cite{schmalensee1989handbook}. Advertising can be viewed as \textit{persuasive}, whereby it changes the tastes of consumers and increases demand (and price) \cite{bloch1999persuasive,crisp1987persuasive,dixit1976advertising};  \textit{informative}, whereby it increases competition and decreases price \cite{telser1964advertising,nelson1975economic,stigler1961economics}; or \textit{complementary}, whereby it appeals to consumers with specific preferences that complement the consumption of the advertised products \cite{becker1993simple,nichols1985advertising,stigler1977gustibus}. These views have drastically different implications. 

In this paper we focus on  \textit{persuasive} advertising  and, as in \cite{braithwaite1928economic}, assume that it increases demand. We look to work by Abernethy and Butler \cite{abernethy1992advertising} to justify this assumption, where they report that an average TV ad contains just one mention of descriptive information about the displayed product (e.g., price, quality, performance, etc.), and that 37.5 percent contain no descriptive information at all. We take this to mean that a significant portion of TV ads are not informative, implying that they are persuasive or complementary. Additionally, we make the simplifying assumption that persuasive advertising is always complementary, as Lindst{\"a}dt and Budzinski argue that the viewer relates with the images and messages for both complimentary and persuasive advertising  \cite{lindstadt2011newspaper}.

A large amount of research has been devoted to using game theory to choose the optimal advertising expenditure to maximize profit \cite{crettez2018existence,marinelli2008optimal,simon1982adpuls,sethi1983deterministic,friedman1983advertising, erickson2011differential, fershtman1984goodwill}. Often this work focuses on settings where there is a monopoly (only one supplier of a good or service) or an oligopoly (only a small number of suppliers of a good or service) \cite{esteban2001informative,erickson2011differential,sethi1983deterministic,friedman1983advertising, fershtman1984goodwill,marinelli2008optimal,gori2015nonlinear}. 

Less research has focused on \textit{monopolistic competition}, where there are many suppliers of a product or service, but the products or services are differentiated only by brand and/or quality. In this paper we develop a model for this setting, looking at the expected advertising expenditure distribution for an arbitrary number of firms competing in a single commodity-product sector.  Our goal is to develop a qualitative understanding of the expected shape of the advertising distribution in a monopolistic competitive setting.

\subsection{Synopsis of modeling approach}

In developing our model, we make the following simplifying assumptions:
\begin{enumerate}
    \item Companies\footnote{We use the terms ``companies'' and ``firms'' interchangeably.} sell an indistinguishable product (except for brand label).
    
    \item There is a linear relationship between the amount of a company's product demanded by the public and the price of the product.
    
    \item Demand for a company's product increases when its advertising is above the mean advertising level and decreases when its advertising is below the mean.

    \item Each company sets the price at a level that maximizes its profit.

    \item Companies continuously adjust their advertising so as to maximize profit.
\end{enumerate}
These assumptions lead to a system of ordinary differential equations describing the dynamics of advertising investments for $N$ firms. 

\begin{figure}[t!]
    \centering
    \includegraphics[width=0.8\columnwidth]{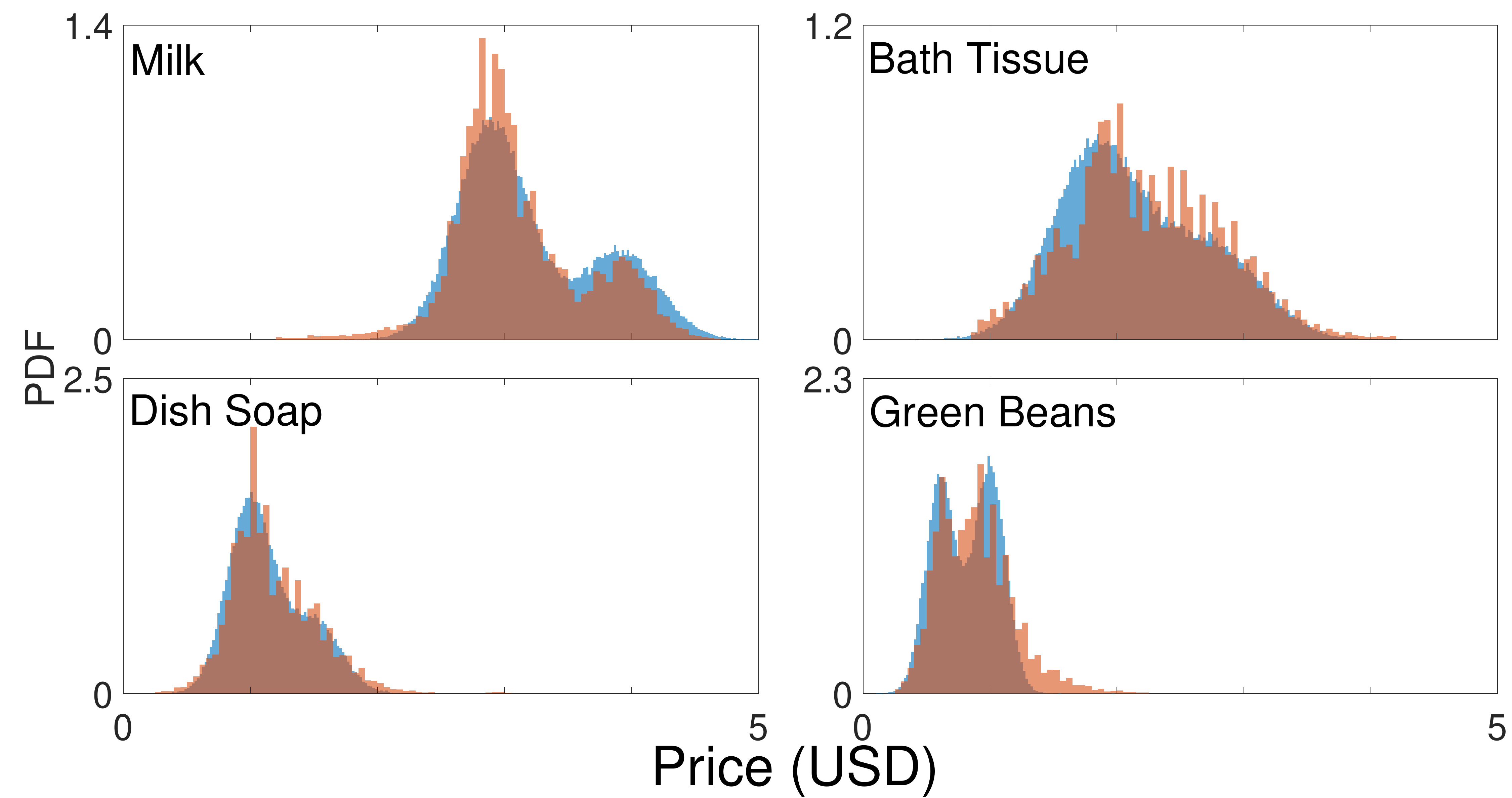}
    \caption{\textbf{Price distributions}.  Red histograms show distribution of prices paid for four common household products; blue histograms show best-fit model predictions.  Data are from a Nielsen database \cite{Kilts2014Nielsen} of over 64 million transactions (purchasing history for 60,000 households).}
    \label{fig:PriceCollage}
\end{figure}

These equations imply that, when advertising is relatively cheap compared to the benefit of advertising, two groups arise: a ``generic brand'' group that advertises a minimal amount, and a ``name brand'' group that advertises at a significantly higher level.\footnote{Note that the minimal advertising level may not be zero: there may be some fixed advertising costs, e.g., associated with product packaging or distribution.  We will treat the minimal level as zero (representing zero ``excess'' advertising) for simplicity in presenting our model, but including an additive constant does not change our predictions.} We find that this segmentation is stable and only ceases to exist when the marginal cost of advertising becomes too high relative to the marginal benefit of advertising. Although our model is intended chiefly to provide a conceptual ``toy'' description, fits to real-world price data\footnote{Data in Fig.~\ref{fig:PriceCollage} have been treated to compensate for psychological pricing, where prices tend to end in certain digits such as "0","5", and "9". See Supplementary Material for details. } show good qualitative agreement (see Fig.~\ref{fig:PriceCollage}).

We caution the reader that, though we use the terms ``generic'' and ``name brand'' to refer to low and high advertising investment states, this is an oversimplification.  In the real world, some brands considered generic may in fact spend significantly on advertising, and some well known ``name brands'' may invest very little in it.  However, for simplicity of exposition, we will employ these terms throughout the manuscript.

\subsection{Outline}

In Section 2, we present our mathematical model for how firms work to differentiate themselves through advertising.  We also present our results relating to existence and stability of equilibria.  In Section 3, we report briefly on the results of numerical experiments to verify consistency with model predictions.  In Section 4, we present real-world data on price distributions for a variety of products and evaluate model fits to assess the consistency of our predictions.  Finally, in Section 5 we discuss other possible applications of our model and limitations of our results.

\section{Model and Analysis} \label{sec:Model}

\subsection{Model Derivation}

Consider $N$ companies (or firms) in a market all selling the same indistinguishable\footnote{By ``indistinguishable'' we mean that the product without branding is indistinguishable, but the brand label is always known to the consumer.}  product. The $i^{\textrm{th}}$ firm purchases a quantity of advertising $a_i$\footnote{This could be quantified, e.g., by clicks on a website ad banner, inserts in a newspaper, views of an ad on TV, or supermarket placement costs.}. For simplicity we assume that the firms have linear demand curves of the form:
\begin{equation}
    Q_i = \Qfree - \kP P_i \;, \qquad i = 1,2, \ldots,N
    \label{demand}
\end{equation}
where $Q_i$ is the quantity demanded of firm $i$'s product, $P_i$ is the unit price for firm $i$'s product, $\Qfree$ is the quantity demanded when the unit price is zero, which may depend on the full distribution of advertising in the market $\vec{a} = (a_1, a_2, \ldots, a_N)$, and $\kP$ is a constant that sets the market's sensitivity to price. 

One measure of a firm's health is the profit generated, with profit defined here as revenue minus production and advertising costs. We take revenue $R_i$ for the $i$th firm to be solely due to sales of this single product at market price:
\begin{equation}
    R_i = Q_i P_i = \Qfree P_i -\kP P_i^2 \ \ \ \ i = 1,2, \ldots,N.
    \label{rev}
\end{equation}

In this model we only consider two types of operating costs: the cost of production $C_{\rmQ}(Q_i)$ and the cost of advertising $C_{\rma}(a_i)$, and we assume an additive relationship 
\begin{equation}
    C_i(Q_i, a_i) = C_{\rmQ}(Q_i) + C_{\rma}(a_i),
    \label{eq:cost}
\end{equation}
where $C_i(Q_i, a_i)$ is the net operating cost for the $i^{th}$ firm. We assume that both $C_{\rmQ}$ and $C_{\rma}$ are increasing functions of their arguments, and for simplicity\footnote{Power laws are common in both natural and engineered systems\cite{markovic2014power,clauset2009power}, and there is evidence that production costs can indeed be approximated by power law scaling \cite{tribe1986scale} .} assume a power law form for each:
\begin{subequations}
\begin{align}
    C_{\rmQ}(Q_i) = \kQ Q_i^{\mu} \label{eq:costq} \\
    C_{\rma}(a_i) = \ka a_i^{\nu} \label{eq:costa} \, ,
\end{align}
\end{subequations}
where $\mu, \nu > 0$ and $\kQ$, $\ka$ are scale factors and can be interpreted as the marginal costs of production and advertising respectively when $\mu =\nu =1$. Thus, the profit function for the $i^{\textrm{th}}$ firm is
\begin{align}
    \pi_i = R_i - C_i = \Qfree P_i - \kP P_i^2- \kQ Q_i^{\mu} -\ka a_i^{\nu}.
    \label{eq:profit}
\end{align}

\begin{figure}[t!]
\centering
\begin{subfigure}{0.45\columnwidth}
    \begin{tikzpicture}
    \draw[->,ultra thick] (0,0) -- (4,0) node[right] {$P_i$}; 
    \draw[->,ultra thick] (0,0) -- (0,3.5) node[above] {$Q_i$};
    \draw[black,ultra thick] +(0,2) -- +(2.33,0) node[yshift = 2cm, left = 2.33cm] {$a_i = \abar$};
    \draw[red,ultra thick,loosely dashed] +(0,3) -- +(3.5,0) node[yshift=3cm, left=3.5cm] {$a_i>\abar$};
    \draw[blue,ultra thick, dotted] +(0,1) -- +(1.16,0)  node[yshift= 1cm,left=1.16cm] {$a_i<\abar$};
    \draw[->,ultra thick] +(1.076,1.25) -- +(1.426,1.66);
    \draw[->,ultra thick] +(0.9,1.05)-- +(0.55,0.64);
    \draw[] (4,4) node{(a)};
    \draw[] (1.16,-0.2) node{\textcolor{white}{$\width$}};
    \draw[] (3.5,-0.2) node{\textcolor{white}{$\width$}};
    \end{tikzpicture}
\end{subfigure}
~ \hspace{-0.25cm}
\begin{subfigure}{0.45\columnwidth}
    \begin{tikzpicture}
    \draw[<->,ultra thick] (-2,0) -- (2,0) node[right] {$a_i-\abar$}; 
    \draw[->,ultra thick] (0,0) -- (0,3.5) node[above] {$\Qfree$};
    \draw[<-,blue,ultra thick] (-2,1) -- (-1,1);
    \draw[blue,ultra thick] (-1,1) -- (0,2);
    \draw[red,ultra thick] (0,2) -- (1,3);
    \draw[->,red,ultra thick] (1,3) -- (2,3);
    \draw[dashed,black,ultra thick] (1,3) -- (0,3) node[left] {$\Qmin+\kAD$} ;
    \draw[dashed,black,ultra thick] (-1,1) -- (0,1) node[right] {$\Qmin$}; 
    \draw[dashed,black,ultra thick] (1,3) -- (1,0)  ;
    \draw[dashed,black,ultra thick] (-1,1) -- (-1,0)  ;
    \draw[] (2,4) node{(b)};
    \draw[] (-1,-0.2) node{$-\width$};
    \draw[] (1,-0.2) node{$\width$};
    \end{tikzpicture}
\end{subfigure}
  \caption{\small{\textbf{Effect of advertising on a firm's demand curve.} (a) Demand shifts due to advertising above (red dashed) or below (blue dotted) the mean level (black solid). Vertical-axis intercepts are $\Qfree$. (b) A simple piecewise linear form for $\Qfree$, the quantity demanded at zero price, which we take to be a non-decreasing function of $a_i - \abar$ that saturates at both left and right limits.  Here the minimum demand (with advertising far below the mean) is $\Qmin$, the maximum demand increase due to advertising is $\kAD$, and the advertising needed beyond the mean for saturation is $\lambda$.} } 
  \label{fig:AdvertEffect}
\end{figure}
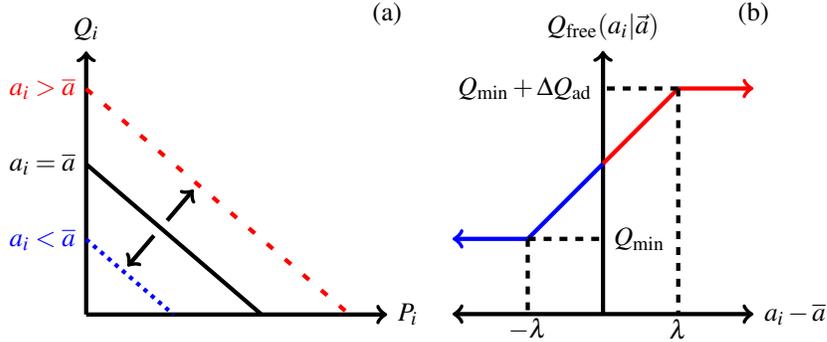

Critically, we tie a firm's level of advertising, $a_i$, to its ability to capture market power. We do this by assuming $\Qfree$ to be a non-decreasing function of $a_i$ referenced to the mean advertising level $\abar = N^{-1} \Sigma_{i=1}^N a_i$, i.e., a non-decreasing function of $a_i-\abar$ (in the most general case, however, it might be an arbitrary function of the full advertising distribution $\vec{a} =(a_1,\ldots,a_N)$). We assume firms that advertise more than the average firm have their demand curves shift out (i.e., quantity demanded increases by a constant amount for all $P_i$) and firms that advertise less than average have their demand curves shift in (i.e., quantity demanded decreases by a constant amount for all $P_i$)---see Fig.~\ref{fig:AdvertEffect}.

We also assume there is a saturation to the amount advertising can influence a firm's ability to capture market share. A plausible smooth, nondecreasing function that saturates is the sigmoid.  We present results for that case in the Supplementary Material (SM).  For greater algebraic simplicity, we define $\Qfree$ here as the following saturating piecewise linear function:
\begin{equation}
    \Qfree = \displaystyle 
    \begin{cases}
        \Qmin, &  a_i -\abar \leq -\width \\
        \Qmin + \dfrac{\kAD}{2\width}  \left(a_i-\abar \right) + \dfrac{\kAD}{2}  , &  -\width < a_i -\abar \leq \width \\
        \Qmin+\kAD, &  a_i-\abar > \width \;,
    \end{cases}
    \label{eq:linkad}
\end{equation}
where $\kAD$ is the maximum demand increase due to advertising, $\Qmin$ is the zero-advertising (minimum) quantity demanded at zero price, which we deem ``intrinsic demand,'' and $\width$ is the width of $\Qfree$ (roughly the amount of excess advertising---above or below the mean---needed for benefits to saturate).  See \FigMid \ref{fig:AdvertEffect} for an illustration. Note, however, that for the purpose of comparison with data, we use the more plausible sigmoidal form:
\begin{equation}
        \Qfree =\dfrac{\kAD}{2} \left\{ \tanh  \left[ \dfrac{a_i -\abar}{\width}\right]+1\right\} + \Qmin.\label{eq:advertresp}
\end{equation}

We assume that each firm always chooses the price $P_i^*$ that maximizes its profit, with corresponding quantity demanded $Q_i^*$. We introduce dynamics to the model by assuming that firms change their advertising levels at a rate proportional to the amount of profit to be gained, i.e,
\begin{equation}
    \tau \dfrac{da_i}{dt} = \frac{\partial \pi_i}{ \partial a_i} = \frac{\partial}{ \partial a_i} %
  \left\{ \Qfree \Pstar(a_i \vert \vec{a}) - \kP [\Pstar(a_i \vert \vec{a})]^2 
  -  \kQ \Qstar(a_i \vert \vec{a})^{\mu} -\ka a_i^{\nu}  \right\}, 
  \label{eq:dadt}
\end{equation}
where the constant $\tau$ sets the time scale for equilibration; we will henceforth take $\tau = 1$ (equivalent to rescaling the time axis) without loss of generality. The list of model parameters with definitions is given in Table \ref{tab:param}.

\begin{table}[t!]
\centering
\begin{tabular}{cc}
    \hline
    Parameter   & Description \\ 
                \hline \hline
    $N$         & Number of companies \\ 
                \hline
    $\Qmin$     & The quantity demanded with minimal advertising at zero price \\ 
                    \hline
    $\kAD$     &  The maximum demand increase due to advertising \\ 
                \hline
    $\kP$       & Decrease in quantity demanded per dollar in unit price increase \\
                \hline
    $\kQ$       & Scale factor for production cost; \\ 
                & (cost of producing an additional unit when costs are linear) \\
                \hline
    $\ka$       & Scale factor for advertising cost; \\ 
                & (cost of producing an additional advertisement when costs are linear) \\
                \hline
    $\width$    & Amount of excess advertising above/below \\ 
                & the mean to achieve maximum/minimum advertising benefits \\
                \hline
    $\mu$       & Scaling exponent in the production cost function \\ 
                \hline
    $\nu$       & Scaling exponent in the advertising cost function \\ 
                \hline
\end{tabular}
\caption{\textbf{Parameter definitions.} Table of parameters used in the model with descriptions.}
\label{tab:param}
\end{table}

\subsection{A concrete example}

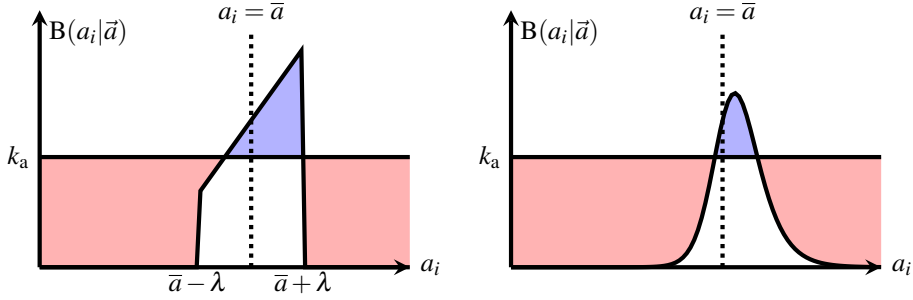
\begin{figure}[t!]
\centering
    { 
    \begin{subfigure}{60mm}
    \begin{tikzpicture}[
    declare function={
        func(\x)= (\x < 1.5) * (0) +
              and(\x >= 1.5, \x < 2.5) *(2*(\x-1.5)+1) +
              (\x >= 2.5) * (0)   ;
    }]
    \begin{axis}[width = 65mm, height = 50mm, 
        xmin=0, xmax=3.5,
        ymin=0, ymax=3.5,
        axis lines=center,
        axis line style = ultra thick,
        axis on top=true,
        clip =false,
        domain=-3.5:3.5,
        ylabel=$\MB(a_i \vert \vec{a})$,
        yticklabels={,,},
        xticklabels={,,},
        tickwidth=0mm
    ]
        \addplot [name path=F, samples=2, black, ultra thick, domain =0:3.5]{1.5};
        \addplot [name path=ka, samples=100, black, ultra thick, domain =0:3.5]{func(x)};
        \addplot fill between[ 
            of = F and ka,
            split, 
            every even segment/.style = {red!30},
            every odd segment/.style  = {blue!30}];
        \addplot +[mark=none,black,ultra thick,dotted] coordinates {(2, 0) (2, 3.2)};
        \draw[] (axis cs: 1.5,-0.2) circle (0pt) node{$\abar-\width$};
        \draw[] (axis cs:2.5,-0.2) circle (0pt) node{$\abar+\width$};
        \draw [black, fill] (axis cs:2,3.2) circle (0pt) node [above] {$a_i =\abar$};
        \draw[] (-0.2,1.5) node{$\ka$};
        \draw[] (3.7,0) node{$a_i$};
    \end{axis}
    \end{tikzpicture}
    \end{subfigure}
    }
    { 
    \begin{subfigure}{60mm}
    \begin{tikzpicture}
    \begin{axis}[width = 65mm, height = 50mm, 
        xmin=0, xmax=3.5,
        ymin=0, ymax=3.5,
        axis lines=center,
        clip =false,
        axis line style = ultra thick,
        axis on top=true,
        domain=-3.5:3.5,
        ylabel=$\MB(a_i \vert \vec{a})$,
        yticklabels={,,},
        xticklabels={,,},
        tickwidth=0mm
    ]
        \addplot [name path=F,samples=2,black, ultra thick, domain =0:3.5]{1.5};
        \addplot [name path=ka, samples=100,black, ultra thick, domain = 0:3.5]{ 2*(tanh(3*(\x-2))+1)*(1-tanh(3*(\x-2))*tanh(3*(\x-2))) };
        \addplot fill between[ 
            of = F and ka,
            split, 
            every even segment/.style = {red!30},
            every odd segment/.style  = {blue!30}];
        \addplot +[mark=none,black,ultra thick,dotted] coordinates {(2, 0) (2, 3.2)};
        \draw [black, fill] (axis cs:2,3.2) circle (0pt) node [above] {$a_i =\abar$};
        \draw[] (axis cs:2.5,-0.2) circle (0pt) node{\textcolor{white}{$\abar+\width$}};    
        \draw[] (-0.2,1.5) node{$\ka$};
        \draw[] (3.7,0) node{$a_i$};
    \end{axis}
    \end{tikzpicture}
    \end{subfigure}
    }
    \caption{\textbf{Advertising dynamics given $\Qfree$.} We plot the the horizontal line $y = \ka$ together with $y = \MB(a_i \vert \vec{a})$ (see \EqMid \eqref{eq:FA}) and add color to indicate the direction of change in advertising according to Eq.~\eqref{eq:dadt}. blue regions show where a firm advertising an amount $a_i$ would choose to increase its advertising and red regions show where a firm advertising an amount $a_i$ would choose to decrease its advertising. Panel (a) shows the case where $\Qfree$ is a piecewise linear function that levels off; (b) shows the case where $\Qfree$ is sigmoidal.}
    \label{fig:adv-dynamics}
\end{figure}

As an analytically tractable example, we first consider the case where production and advertising costs grow at a linear rate, i.e, $\mu = \nu = 1$. Substituting Eq.~\eqref{demand} into  Eq.~\eqref{eq:profit}, setting $[\partial \pi_i / \partial P_i]_{P_i=P_i^*} = 0$ and solving for the profit-maximizing price $P_i^*$ gives
\begin{equation}
    P_i^*(a_i \vert \vec{a}) = \frac{1}{2}\left[\Qfree/\kP + \kQ \right]. \label{eq:Pstar}
\end{equation}
The corresponding profit-maximizing quantity is 
\begin{equation}
    Q_i^*(a_i \vert \vec{a}) = \frac{1}{2}\left[\Qfree -\kQ \kP \right]. \label{eq:Qstar}
\end{equation}
Substituting this into Eq.~\eqref{eq:dadt} yields the dynamical system
\begin{equation}
    \dfrac{da_i}{dt} = \MB(a_i\vert \vec{a})- \ka, \label{eq:dadt_gen}
\end{equation}
where $\MB(a_i \vert \vec{a})$ is defined as
\begin{equation}
       \MB(a_i \vert \vec{a}) = \displaystyle \begin{cases}
      \dfrac{N-1}{N}\dfrac{\kAD}{4\width\kP}\left[\dfrac{\kAD}{2\width} \left(a_i-\abar \right) + \dfrac{\kAD}{2}+\Qmin-\kQ \kP  \right] , &  \vert a_i -\abar \vert < \width \\
        0, &  \vert a_i -\abar \vert > \lambda
        \end{cases}. \label{eq:FA}
\end{equation}
$\MB(a_i \vert \vec{a})$ represents the marginal benefit of advertising and $\ka$ the marginal cost of advertising. For any firm with advertising close enough to the mean ($|a_i-\abar| < \lambda$), the function $\MB$ is simply a line of positive slope $(N-1)\kAD^2/(8 N \width^2\kP) \xrightarrow[N \to \infty]{} \kAD^2 / 8 \width^2 \kP$.  Firms with $\MB > \ka$ have $d a_i/d t > 0$ and increase their advertising budgets, while firms with $\MB < \ka$  decrease their advertising budgets.  For all firms far from the mean ($|a_i-\abar| > \lambda$), $\MB = 0$ and thus $d a_i/d t = -\ka < 0$. This flow is illustrated in the left panel of \FigMid\ref{fig:adv-dynamics}.  The corresponding flow in the case of a smooth sigmoidal $\Qfree$ is shown in the right panel of the same figure.  The intuition drawn from the piecewise case outlined here applies similarly to the sigmoid.

\subsubsection{Existence of equilibria} \label{sec:exist}

For a given $\abar$, there can be at most three fixed points. In \FigMid \ref{fig:adv-dynamics}, three fixed points are located at $a_i^* = \abar+ \width$, the intersection where $\MB(a_i^* \vert \abar) = \ka$ for $a_i^* < \abar$, and at $a_i^* = 0$  (since advertising cannot be negative). Because stability must alternate for one-dimensional flows, $a_i^* = 0$ and $a_i^* = \abar + \width$ are the only stable fixed points. Thus, any stable equilibrium distribution $\vec{a}$ with mean $\abar$ must have $a_i = 0$ or $a_i = \abar+\width$ for all $i$.

We refer to the case when advertising is bimodal as the \textit{differentiated} state.  We note that such a state may only exist when two stable fixed points exist, which requires $\max_{a} \MB(a\vert\vec{a}) > \ka$. Letting $N\rightarrow\infty$, one can write this condition explicitly as
\begin{equation}
    \max_{a_i} \dfrac{\partial\pi_i}{\partial a_i}= \dfrac{\kAD\left(\kAD+ \Qmin- \kQ \kP \right)}{4\width\kP} -\ka >0. \label{eq:MaxF}
\end{equation}
Put simply, if advertising does not increase profit anywhere, bimodality cannot arise. 

The differentiated state must be self-consistent---that is, the two stable fixed points (at $0$ and $\abar+\lambda$) when averaged (including weights based on corresponding firm fractions) must yield the appropriate mean advertising $\abar$. We refer to the fraction of firms that choose to set their advertising to zero (``generics'') as $x$, and thus the fraction of firms that set their advertising to $a_i^* = \abar+\lambda$ (``name brands'') is $1-x$. It follows then that $\abar = 0 x + a_i^* (1-x) = (\abar + \lambda)(1-x)$ must hold for self-consistency. Solving for $\abar$ gives
\begin{equation}
    \abar = \frac{\lambda (1-x)}{x}~,
    \label{eq:abar}
\end{equation}
and the name-brand advertising level at equilibrium is then $a_i^* = \abar + \lambda = \lambda/x$.

We also require that the unstable fixed point at $\MB(a_i^* \vert \abar) = \ka$ be non-negative, assuming the system has reached self-consistent equilibrium (i.e., $\abar =\lambda (1-x)/x$ as in Eq.~\eqref{eq:abar}). If the unstable fixed point were negative, that would imply $da_i/dt > 0$ over the entire domain $0 \leq a_i < \lambda/x$, which would contradict the assumption of a stable fixed point at $a_i=0$ that went into the self-consistency argument above. 

Imposing this constraint on the unstable fixed point defined by $\MB(a_i^* \vert \abar) = \ka$, we find
\begin{equation}
    a_i^{*\textrm{(unstable)}} = \frac{\lambda}{x\kAD^2 }\left[\kAD^2 (1-2x) - 2x\kAD (\Qmin-\kP\kQ) +8x\lambda \left(\frac{N}{N-1}\right)\ka \kP  \right]
\end{equation}
and
\begin{equation}
    \xcrit =  \frac{\kAD^2/2}{\kAD^2+ \kAD  (\Qmin-\kP\kQ) - 4\lambda\left(\frac{N}{N-1}\right)\kP \ka }. 
    \label{eq:xcrit}
\end{equation}

Here, $\xcrit$ bounds the feasible proportion of generic firms from above. Equations $\eqref{eq:MaxF}$ and \eqref{eq:xcrit} (with $x < \xcrit$) establish necessary conditions for existence of the differentiated state. It is not feasible to derive an equivalent analytical expression for $\xcrit$ in the case of sigmoidal $\Qfree$, but it is straightforward to numerically compute $\xcrit$ for a given set of parameters, as we do in the SM.

Another state is possible where all firms set their advertising to zero---we refer to this as the \textit{undifferentiated} state. Clearly from \EqMid \eqref{eq:dadt_gen}, $\max_{a} \MB(a\vert\vec{a}) < \ka$ implies that $da_i/dt <0$ for all $a_i$. In this case, $a_i^* =0$ for all $i$ is the only equilibrium.

\subsubsection{Stability of equilibria} \label{sec:stabeq}

We now consider the stability of the differentiated and undifferentiated states. First, we focus on the stability of the differentiated state. We assume there exists an equilibrium with $Nx$ ``generic'' firms choosing to invest nothing in advertising, and $N(1-x)$ ``name-brand'' firms choosing to advertise at level $\abar+\width$, with $0<x<1$ representing the proportion of ``generic'' firms. Assuming that $N \gg 1$\footnote{Numerical experiments suggest that stability conditions derived in this section also hold for small $N$.} and hence that a small perturbation of a single firm has a negligible impact on the mean $\abar$, we consider perturbation of the $i$th ``name-brand'' firm's advertising by an amount $\delta$ and track how $\delta(t)$ changes in time. That is, we set $a_i = \abar + \width  + \delta(t)$, which yields the system
\begin{equation}
    \frac{d\delta}{dt} = \displaystyle 
    \begin{cases}
        \dfrac{N-1}{N}\dfrac{\kAD}{4\width\kP}\left[\dfrac{\kAD}{2\width} \delta + \kAD+\Qmin-\kQ \kP \right]-\ka , &  \vert \delta +\width \vert < \width \\
        -\ka, &  \vert \delta + \width \vert > \width
    \end{cases}.
    \label{eq:linEq_name}
\end{equation}
If the condition for existence of the differentiated state given in \EqMid \eqref{eq:MaxF} holds, sufficiently small $|\delta|$ implies that $d\delta/dt >0$ when $\delta <0$. Additionally, it is clear that $d\delta/dt <0$ when $\delta >0$. Thus, under this type of perturbation the differentiated state is stable. If we similarly perturb one firm from the generic group, i.e., setting $a_i = \delta >0$, we find
\begin{equation}
    \dfrac{d\delta}{dt} = \displaystyle
    \begin{cases}
        \dfrac{N-1}{N}\dfrac{\kAD}{4\width \kP}\left[\dfrac{\kAD}{2\width} \left(\delta-\abar \right) + \dfrac{\kAD}{2}+\Qmin-\kQ \kP  \right]-\ka , &  \vert \delta -\abar \vert < \width \\
        -\ka, &  \vert\delta-\abar\vert > \width
    \end{cases}.
    \label{eq:linEq_gen}
\end{equation}
If $\abar > \width$ then there exists $\delta>0$ small enough such that $d\delta/dt < 0$ since $\delta < \abar +\width$ implies that $d\delta/dt=-\ka <0$. If $\abar < \width$ then $d\delta/dt$ is given by the linear equation in \EqMid \eqref{eq:linEq_gen} for small $\delta$. Thus, the differentiated state is stable under such a perturbation when
\begin{equation}
    \left.  \dfrac{d\delta}{dt}\right\vert_{\delta\rightarrow 0^+} =      \dfrac{N-1}{N}\dfrac{\kAD}{4\width \kP}\left[-\dfrac{\kAD}{2\width} \abar  + \dfrac{\kAD}{2}+\Qmin-\kQ \kP  \right]-\ka  = \left.\dfrac{\partial \pi_i}{\partial a_i}\right\vert_{a_i =0} < 0. 
\end{equation}
This means it must be unprofitable for companies with no advertising to increase their advertising for the differentiated state to be stable.

Now we consider the stability of the undifferentiated state, $a_i =0$ for all $i$. As stated in Section \ref{sec:exist}, if $\max_{a} \MB(a\vert\vec{a}) < \ka$, then $da_i/dt <0$ for all values of $a_i$. Thus, it is clear that the undifferentiated state exists and is stable in that case. We now focus on the case where $\max_{a} \MB(a\vert\vec{a}) > \ka$. If $a_i = 0$ for all $i$ then $\abar =0$. We consider a perturbation of one firm from this state. Letting $a_i=\delta$ again we get
\begin{equation}
    \dfrac{d\delta}{dt} = \displaystyle
    \begin{cases}
        \dfrac{N-1}{N}\dfrac{\kAD}{4\width \kP}\left[\dfrac{\kAD}{2\width} \left(\delta \right) + \dfrac{\kAD}{2}+\Qmin-\kQ \kP \right]-\ka , &  0 \leq \delta  < \width \\
        -\ka, &  \delta > \width
    \end{cases}.
    \label{eq:linEq_zero}
\end{equation}
If $d\delta/dt < 0$ when $\delta = 0$ by continuity of the $d\delta/dt$ in the range of $0 < \delta < \lambda$ there must exist some $\delta > 0$ sufficiently small such that $d\delta/dt < 0$. Therefore, the system is stable under this kind of perturbation when
\begin{equation}
    \left.  \dfrac{d\delta}{dt}\right\vert_{\delta\rightarrow 0} = \dfrac{N-1}{N}\dfrac{\kAD [\Qmin+\kAD/2-\kQ \kP] }{4 \width \kP} - \ka = \left.\dfrac{\partial \pi_i}{\partial a_i}\right\vert_{a_i =\abar} <0.
    \label{eq:UD_stab}
\end{equation}
We surmise from this that the undifferentiated state is stable only if increasing advertising is not profitable for the average firm. 

The above stability arguments can be generalized to arbitrary infinitesimal perturbations of the advertising distribution in the limit $N \to \infty$. See work by Clifton, Braun, and Abrams for a description of such an approach in a different context \cite{clifton2016handicap}.

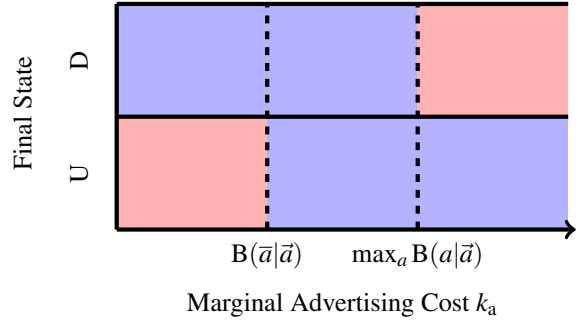
\begin{figure}[t!]
    \centering
    \begin{tikzpicture}[scale=0.5] 
        \fill [red!30] (0,0) rectangle (4,3);
        \fill [blue!30] (4,0) rectangle (8,3);
        \fill [blue!30] (8,0) rectangle (12,3);
        \fill [blue!30] (4,6) rectangle (0,3);
        \fill [blue!30] (8,6) rectangle (4,3);
        \fill [red!30] (12,6) rectangle (8,3);
        \draw[-,ultra thick] (0,3) -- (12,3) node{};
        \draw[-,ultra thick] (0,6) -- (12,6) node{};
        \draw[-,ultra thick,dashed] (4,0) -- (4,6) node{};
        \draw[-,ultra thick,dashed] (8,0) -- (8,6) node{};
        \draw[->,ultra thick] (0,0) -- (12.2,0) node{};
        \draw[-,ultra thick] (0,0) -- (0,6) node{};
        \draw[] (6,-2) node{Marginal Advertising Cost $\ka$};
        \draw[] (4,-0.7) node{$ \MB(\abar \vert \vec{a}) $};
        \draw[] (8,-0.7) node{$\max_a \MB(a \vert \vec{a}) $};
        \draw[] (-1,1.5) node{\rotatebox{90}{{\normalsize  U}}};
        \draw[] (-1,4.5) node{\rotatebox{90}{{\normalsize D}}};
        \draw[] (-2.5,3.1) node{\rotatebox{90}{{\normalsize  Final State}}};
    \end{tikzpicture}
    \caption{\textbf{Regions of stability.} We illustrate the regions of stability for the differentiated and undifferentiated states, indicated in the figure by D and U respectively. These are given by Eqns.~\eqref{eq:MaxF}, \eqref{eq:UD_stab}, and \eqref{eq:BiStable_Norm}. Here blue indicates that a state is stable and red indicates that a state is unstable. The middle column, where $\MB(\abar\vert\vec{a}) < \ka < \max_{a} \MB(a\vert\vec{a})$ is where both states are stable.}
    \label{fig:stability}
\end{figure}

\FigBeg \ref{fig:stability} maps the regions of stability for the differentiated and undifferentiated states given by Eqns.~\eqref{eq:MaxF}, \eqref{eq:UD_stab}, and \eqref{eq:BiStable_Norm}. Both the differentiated and undifferentiated states can be simultaneously stable. If Eqns.~\eqref{eq:MaxF} and \eqref{eq:UD_stab} both hold then both states are stable. Thus, we write the condition for bistability as 
\begin{align}
 \MB(\abar \vert\vec{a}) = \dfrac{N-1}{N}\dfrac{  \kAD [\Qmin+\kAD/2-\kQ \kP] }{4 \width \kP} < \ka <\max_{a} \MB(a\vert\vec{a}).\label{eq:BiStable_Norm}
\end{align}
In more intuitive terms,
\begin{equation}
\left.\dfrac{\partial \pi_i}{\partial a_i}\right\vert_{a_i =\abar}  < 0 < \max_a \dfrac{\partial \pi_i}{\partial a_i}.
\label{eq:intuit_stab}
\end{equation}

Thus, bistability of the differentiated and the undifferentiated states occurs when the maximum marginal profit is positive, but it is profitable for the average firm to decrease its advertising. The regions of stability of the undifferentiated  and differentiated states are defined similarly when $\Qfree$ is sigmoidal (and hence, $\MB(a_i \vert \vec{a})$ altered appropriately---see SM).

\section{Numerical experiments} \label{sec:numexp}

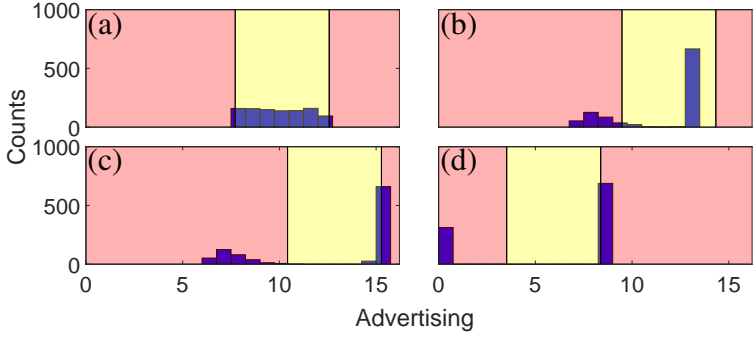
\begin{figure}[t!]
    \centering
    \begin{tikzpicture}
        \node at (0,0) {\includegraphics[width=0.8\columnwidth]{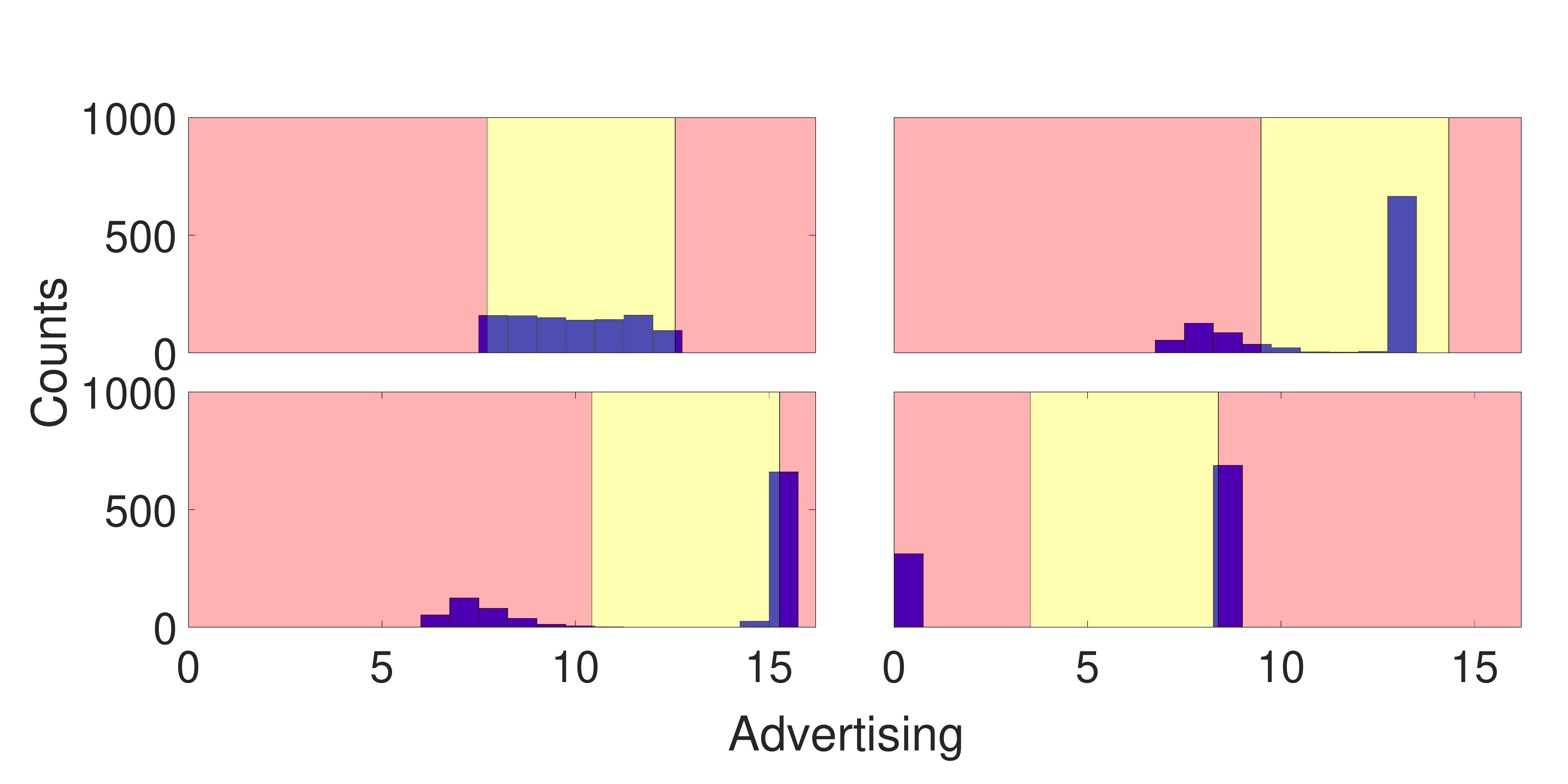}};
        \node at (-3.7,1.6) {{\large(a)} };
        \node at (1,1.6) {{\large(b)}};
        \node at (-3.7,-0.2) {{\large(c)}};
        \node at (1,-0.2) {{\large(d)}};
    \end{tikzpicture}
    \caption{\textbf{Simulation of the system.} In this figure we give snapshots of the numerical integration of the system from initial condition to equilibrium. In panel (a), the system starts from the uniform randomly distributed state with the advertising initial condition set as $\mathcal{U}(7.5,12.5)$. In panel (b), the separation into two groups has begun. Companies change their spending until the lower group is far away from the mean, as seen in panel (c). Finally, in panel (d), a bimodal equilibrium has been reached, with one group representing generic brand companies ($a\approx0$) and the other representing name-brand companies (advertising at a nonzero value at $a=a_{\textrm{name}}$). The green areas indicate where companies will increase their advertising and the red areas indicate areas where companies will decrease their advertising. In this simulation we set the number of companies to $N=1000$, $\ka = \kP = \kQ = \nu= \mu = \width = 1$, and $\Qmin = \kAD = 10$ (see Table \ref{tab:param} for parameter definitions).
    }
    \label{fig:BimodSim}
\end{figure}

In order to test model predictions we perform simple numerical experiments; all results appear to be consistent with theory.  \FigBeg \ref{fig:BimodSim}  shows an example of a simulation where the benefit of advertising saturates (we assume a sigmoidal functional form) when $\ka < \max_{a} \MB(a\vert\vec{a})$. Starting from a uniformly distributed initial condition, the firms arrange themselves so that there is a ``generic'' group at advertising level $a=0$ and a ``name-brand'' group at $a=a_{\textrm{name}}>0$. Colors have been added to indicate ranges where firms decrease (red) or increase (yellow) their advertising (see figure caption for details).

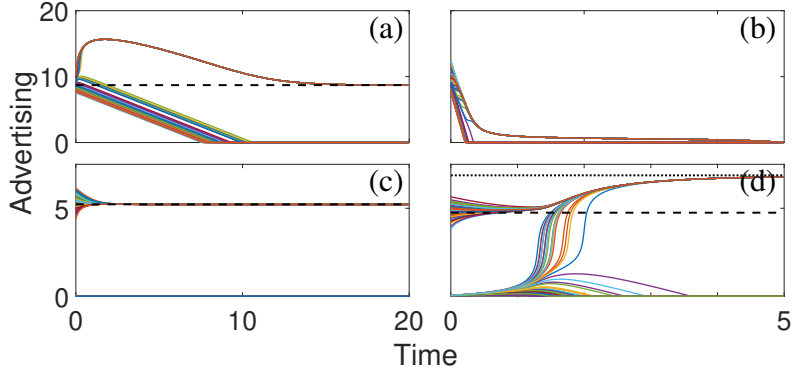
\begin{figure}[t!]
    \centering
    \begin{tikzpicture}
        \node at (0,0) {\includegraphics[width=0.8\columnwidth]{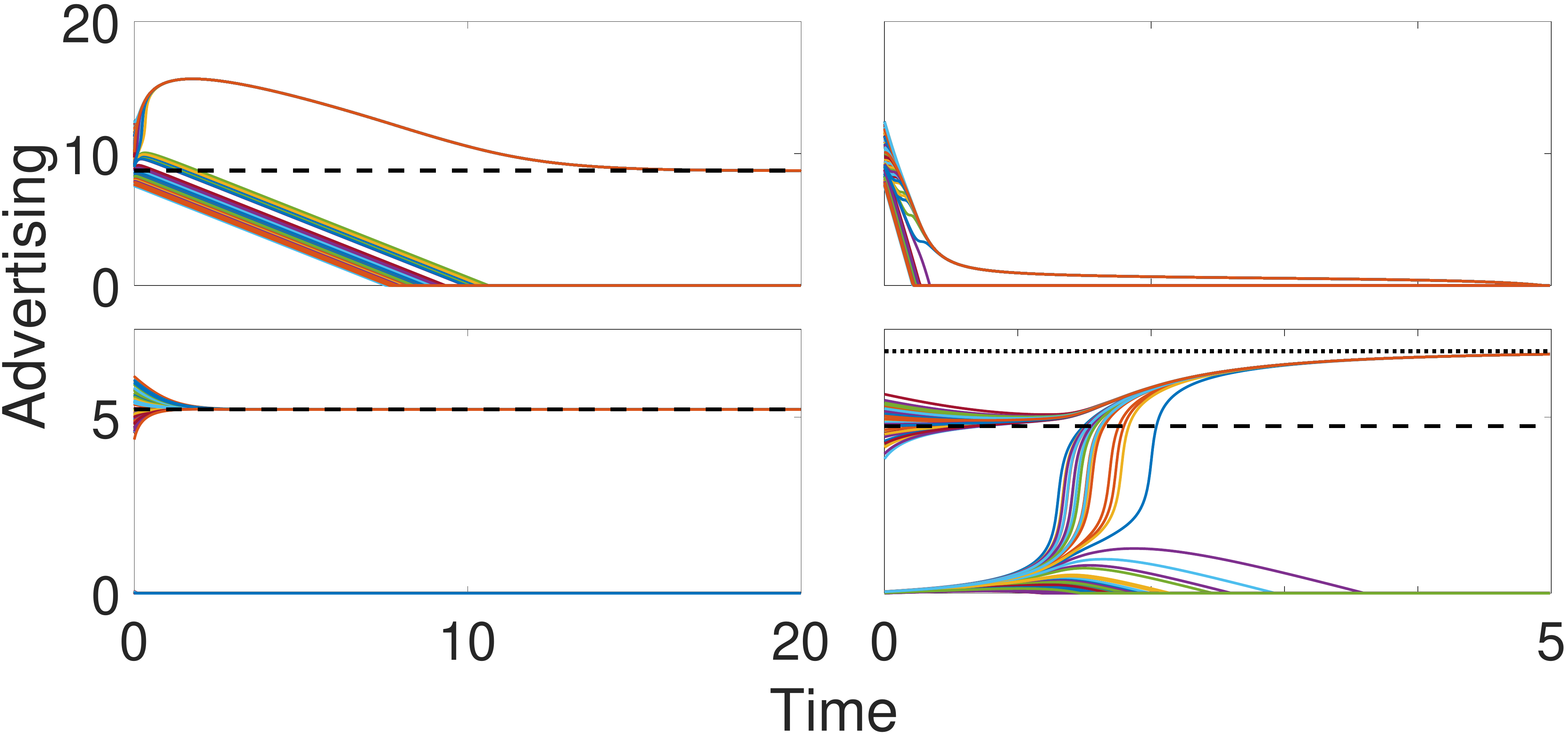}};
        \node at (-0.19,2.04) {{\large(a)} };
        \node at (4.75,2.04) {{\large(b)}};
        \node at (-0.19,0.01) {{\large(c)}};
        \node at (4.75,0.01) {{\large(d)}};
    \end{tikzpicture}
    \caption{\textbf{Numerical exploration}. We illustrate the qualitative states described in Section \ref{sec:Model}. In panels (a) and (b), we set the initial condition by sampling from the uniform random distribution $\mathcal{U}(7.5,12.5)$. In panel (a), $\ka < \max_{a} \MB(a\vert\vec{a})$, and the firms approach into a bimodal distribution (differentiated state).  In panel (b), $\ka > \max_{a} \MB(a\vert\vec{a})$, and the firms approach the zero advertising  (undifferentiated) state. In panels (c) and (d), we set the initial condition by perturbing off of a theoretical differentiated state with $\xcrit = 0.53$.  In panel (c) we set $x = 0.5 < \xcrit$; after perturbation the system returns to the differentiated state. In panel (d) we set $x = 0.55 > \xcrit$; after perturbation some firms move from the generic group to the name-brand group so that the final generic fraction $x = 0.38$ is less than $\xcrit = 0.53$.  Dashed lines show theoretical advertising level for name-brand group before initial perturbation.  Dotted line in panel (d) shows theoretical advertising level for name-brand group after fraction generic has changed to final value.  In all simulations we set $N = 100$, $ \kP = \kQ = \nu= \mu = \width = 1$, and $\Qmin = \kAD = 10$. }
    \label{fig:num_cases}
\end{figure}

Figure \ref{fig:num_cases} demonstrates some of the existence and stability boundaries outlined in Section \ref{sec:Model}. Panels (a) and (b) start with an initial condition that is sampled from the uniform random distribution $\mathcal{U}(7.5,12.5)$. When $\ka < \max_{a} \MB(a\vert\vec{a})$ (panel (a)), firms separate into two groups and move toward the differentiated state. When $\ka > \max_{a} \MB(a\vert\vec{a})$ (panel (b)), all firms tend to zero advertising.

Panels (c) and (d) each begin with bimodal advertising distributions, differing only in the initial fractions generic $x$. In panel (c), initially $x < x_{\textrm{crit}}$, and the system relaxes to a stable differentiated state.  In panel (d), initially $x > x_{\textrm{crit}}$, and, since no nearby differentiated state exists for that fractionation, the system ends up at a different differentiated state (where $x < x_{\textrm{crit}}$) due to some firms transitioning from the generic group to the name-brand group. See figure caption for more detail.

In SM, we discuss simulation of other variants of our model including  nonlinear production cost curves ($\mu \neq 1$) and advertising cost curves ($\nu \neq 1$), and nonidentical firms (e.g., nonuniform $\Qmin$ and/or $\kAD$).  We encountered no qualitative difference in the results for those cases.



\section{Data}

We use price data from the Nielsen Corporation. Nielsen's consumer panel data contains annual shopping information from thousands of American households, starting from 2004 with yearly updates. Individuals involved in the study used in-home scanners to record all of their purchases that were designated for personal use. Scanners recorded each product's Universal Product Code (a string of digits that uniquely identify the product) and the product's price. We analyze data from 2014 containing over 64 million transactions from 60,000 households \cite{Kilts2014Nielsen}. 

Our model's primary prediction is the distribution of advertising investments across firms.  While we would have preferred to employ data that directly reflects such advertising budgets, we were not able to find any source comparable in quality to the Nielsen price dataset.  Nevertheless, our model also carries with it predictions for prices, though we must accept that real-world prices may be additionally influenced by other unmodeled factors.  In working with price data, we make the assumption that these other unmodeled factors either have negligible impact or do not change the unimodal/multimodal nature of the distribution.

\subsection{Fitting procedure} \label{sec:fitting}

To fit our model predictions to data, we first define an objective function $H[f(p), g(p)]$ to quantify the difference between distributions predicted by the model ($f(p)$) and inferred from the data ($g(p)$). Specifically, we set our objective function $H[f(p), g(p)]$ to be the square integrated difference between the distributions
\begin{equation}
  H(f, g) = \int_{-\infty}^{\infty} [f(p) -g(p)]^2  dp.
  \label{eq:ObjFun}
\end{equation}

We use the Nelder-Mead algorithm \cite{nelder1965simplex} to minimize this objective function over a subset of parameters that most directly affect the demand curve given in \EqMid \eqref{demand}: the maximum benefit from advertising $\kAD$, the minimum quantity demanded $\Qmin$, and consumers' price sensitivity, $\kP$.  If the data indicate bimodality, we also optimize over the generic fraction $x$.

We model heterogeneity among firms by adding random variables $\zeta_i$ and $\xi_i$ to the parameters $\kAD$ and $\Qmin$ respectively. These random variables are drawn from a normal distribution with mean zero and respective standard deviations $\epsilon_1$ and $\epsilon_2$. We interpret $\epsilon_1$ as the variation in the quality of advertising messaging and $\epsilon_2$ as the variation in natural demand for the firms' products, and we also optimize their values.

We must choose starting ``seeds'' for the Nelder-Mead algorithm since it is a local optimization method. We do this by first extracting two modes from the price data, one which corresponds to the lower advertising investment group (generics) and the other the higher advertising investment group (name brands).  We then choose seed parameters such that the model’s predicted price distribution matches up with those two modes. For a more detailed description of the initialization of the algorithm see the SM.

\FigBeg \ref{fig:PriceCollage} provides a few examples of fits for products that had more than 10,000 transactions. These examples also demonstrate the variety of products within the dataset. We see there is qualitative agreement between the model's predicted price distributions and the empirical price distributions.

\subsection{Statistics}\label{sec:stat_tests}

We attempt to validate our model by fitting theoretical price distributions to empirical data provided by Nielsen Corporation \cite{Kilts2014Nielsen}. We use two tests, the Kolmogorov-Smirnov (KS) test and Hartigan's Dip Test, to assess the quality of our fits. See Figure \ref{fig:PriceCollage} for a sample of model fits to data.

The KS test generates the probability that two samples come from the same underlying distribution by calculating the maximum absolute difference between their cumulative distribution functions (CDFs). Here, a large difference implies a low probability that the two datasets come from the same distribution.  For a majority ($58\%$) of our model fits to the top 500 products, we fail to reject the null hypothesis (samples from same underlying distribution) at a significance level of $0.05$: the data and the model prediction may come from the same distribution.

Hartigan's Dip Test assesses whether a distribution is unimodal by comparing the CDF of the distribution to a unimodal test distribution \cite{hartigan1985dip}.  A large difference between the distribution in question and the test distribution indicates a low probability of the distribution being unimodal.  We apply Hartigan's Dip Test to the 500 products with the most entries in the database, and find that $46\%$ have price distributions inconsistent with unimodality at a significance level $0.05$.  If price distributions are linked to advertising expenditures, as our model indicates, then almost half the products have a multimodal (bimodal or higher number of modes\footnote{We suspect that an extension of this model to allow stronger within-segment competition (i.e., name brands compete more strongly with each other than with generics) would lead to additional modes.}) advertising distribution.  For other products, unimodality could not be rejected, but data may not be inconsistent with bimodality. See the SM for the full distribution of $p$-values.

\section{Discussion}

The theory we present provides a possible explanation for the segmentation of commodity-product sectors into ``name brand'' and ``generic'' products.  We speculate that similar explanations might exist for other contexts where hierarchy emerges as a result of competition, or where interactions between individual agents can lead to global patterns \cite{page2012aggregation}.  For example, competition for a mate \cite{mccullough2016sexually, clifton2016handicap, kirkpatrick1982sexual} and competition for resources \cite{tilman1977resource, hibbing2010bacterial, auger1998hawk} can both result in hierarchies observed in the natural world.  Our model might be adapted to yield insight into such phenomena.

\subsection{Limitations}

In creating a highly simplified model, we have inevitably made some assumptions that limit its generality. These include: 
\begin{itemize}
	\item We assumed that advertising was persuasive and hence, that quantity demanded increased uniformly across all price levels as advertising increased. In cases where advertising is informative, however, one would expect the slope of the demand curve to increase, instead of simply shifting vertically. 
	
	\item We chose to leave the development of brand loyalty out of our model.  This could presumably be captured through a demand curve that becomes more inelastic as loyalty increases.
	
	\item We excluded spillover effects from ``generic advertising'' whereby advertising leads to increases in demand for all companies selling a similar product \cite{norman2008generic,bass2005generic}. We expect that this would increase profit for all companies but not affect the bimodal segmentation our current model predicts.
	
	\item We assumed advertising has stable and lasting impact.  Our model treats the benefits of advertising as arising instantaneously, an approximation that is only merited when the time scale of interest is much longer than the advertising's ``half-life'' in the consumer environment. 
	
	\item We assumed the existence of many producers selling similar products.  Some of our arguments would not be valid in the case of an oligopoly, where there are only a handful of producers.
	
	\item We approximated demand curves as linear, but of course these could (and likely do) take on more complex forms for real products.  
\end{itemize}

In addition to the limitations of our modeling approach, the data set we examine also contains some biases that should be pointed out.  Most saliently, the price distributions we examine are the result of different vendors selling identical products for different prices: this means that branding is really present at the vendor level, slightly different from the most direct and natural interpretation of the model.  Also, a large fraction of entries in the database are food and other consumable products, since these are purchased more frequently than durable goods.  Consumables might have a different market structure than products in non-food markets (e.g., electronics, health care, housing, etc.).

\subsection{Conclusions}

We have presented a simple mathematical model for competition among firms on the basis of advertising.  Despite the model's simplicity, a surprisingly robust prediction emerges: products split into ``name brand'' and ``generic'' groups. This prediction appears to be largely consistent with data both in a qualitative sense (many products have non-unimodal price distributions) and a quantitative sense (theoretical price distributions from the model are consistent with empirical price distributions), even without a more detailed and accurate model. 

Advertising has a large macroeconomic impact on corporate profits, market efficiency, and consumer welfare.  The segmentation we report contrasts starkly with (often implicit) assumptions of smooth, singly-peaked functions for economic metrics. We hope that our work helps refine intuition and inspires further inquiries into this intriguing aspect of free market dynamics.

\section*{Data Availability}

The data that support the findings of this study are available from The Nielsen Company (US), LLC but restrictions apply to the availability of these data, which were used under license for the current study, and so are not publicly available. Data are however generally available for scientific research with an institutional or individual subscription \cite{Kilts2014Nielsen}.

\section*{Acknowledgements}
The authors gratefully acknowledge NSF support through
Research Training Grant No. 1547394. We thank Nielsen Corporation for their generosity in making their proprietary data available for scientific research.  We also thank Vicky Chuqiao Yang for useful conversations.

\section*{Author Contributions}
DMA and JDJ conceived of the research. JDJ performed the majority of the model analysis and numerical integration/simulation. AMR helped write code for numerical simulation. DMA and JDJ wrote the manuscript together.

\begin{footnotesize}
\textit{Researcher(s) own analyses calculated (or derived) based in part on data from The Nielsen Company (US), LLC and marketing databases provided through the Nielsen Datasets at the Kilts Center for Marketing Data Center at The University of Chicago Booth School of Business.}
\end{footnotesize}

\begin{footnotesize}
\textit{The conclusions drawn from the Nielsen data are those of the researcher(s) and do not reflect the views of Nielsen. Nielsen is not responsible for, had no role in, and was not involved in analyzing and preparing the results reported herein.}
\end{footnotesize}

\bibliographystyle{siamplain}

\bibliography{advert}

\end{document}


\maketitle

\section{Case with sigmoidal advertising payoff} \label{sec:sigmoidal_payoff}

In Section \secModel, we carried out our analysis assuming that the quantity demanded at zero price, $\Qfree$, took the form of a piecewise linear function given by Eq.~\eqlinkad. Here, we assume $\Qfree$ is given by a sigmoid:
\begin{equation}
    \Qfree =\dfrac{\kAD}{2} \left\{ \tanh  \left[ \dfrac{a_i -\abar}{\width}\right]+1\right\} + \Qmin\ .
    \label{eq:advertresp_sig}
\end{equation}
As in Section \secModel, we first consider the case where production and advertising costs scale linearly, i.e, $\mu = \nu = 1$. We substitute Eqns.~\eqPstar, \eqQstar, and \eqref{eq:advertresp_sig} into Eq.~\eqdadt, giving
\begin{align}
    \tau\dfrac{da_i}{dt} &=  \dfrac{N-1}{N} \dfrac{ \kAD}{8\width \kP}  \left[\Qfree -\kQ \kP \right]\sech^2\left\{ \dfrac{a_i -\abar}{\width}\right\} -\ka \nonumber \\
     &= \MB(a_i \vert \vec{a})-\ka, \label{eq:linEq_Sigmoid}
\end{align}
where 
\begin{equation}
    \MB(a_i \vert \vec{a}) = \dfrac{N-1}{N} \dfrac{ \kAD}{8\width \kP}  \left[\Qfree -\kQ \kP \right]\sech^2\left\{ \dfrac{a_i -\abar}{\width}\right\}\;. 
\end{equation}
Similarly to Section \secModel, $\tau$ sets the time scale for equilibration, and we set $\tau =1$ without loss of generality. 

Fixed points occur when $\MB(a_i \vert \vec{a})$ intersects $\ka$. \FigBeg \advdynamics~shows how firms would change their advertising given mean $\abar$, according to Eq.~\eqref{eq:linEq_Sigmoid}. Of course, since $\abar$ changes in time due to the distribution of advertising $\vec{a}$ evolving in time, $\MB(a_i \vert \vec{a})$ is variable in time as well. For a given $\abar$\footnote{But of course $\abar$ is itself a dynamical variable, and a self-consistent solution must be sought where the equilibrium distribution of firm advertising implied by $\abar$ is consistent with $\abar$.} there are three fixed points for the system: two from the intersection of $\MB(a_i \vert \vec{a})$ with $\ka$ and one implied at zero as negative advertising is not physically meaningful in this problem. One can see from \FigMid \advdynamics~that, given $\abar$, the equilibrium would involve a distribution with some companies at zero and some at $a= a_{\textrm{name}} >0$. Furthermore, the equilibrium must be distributed such that $\abar = 0 x + a_{\textrm{name}} (1-x)$, where $0<x<1$ gives the fraction of firms generic. 

\begin{figure}[t!]
    \centering
    \includegraphics[width=110mm]{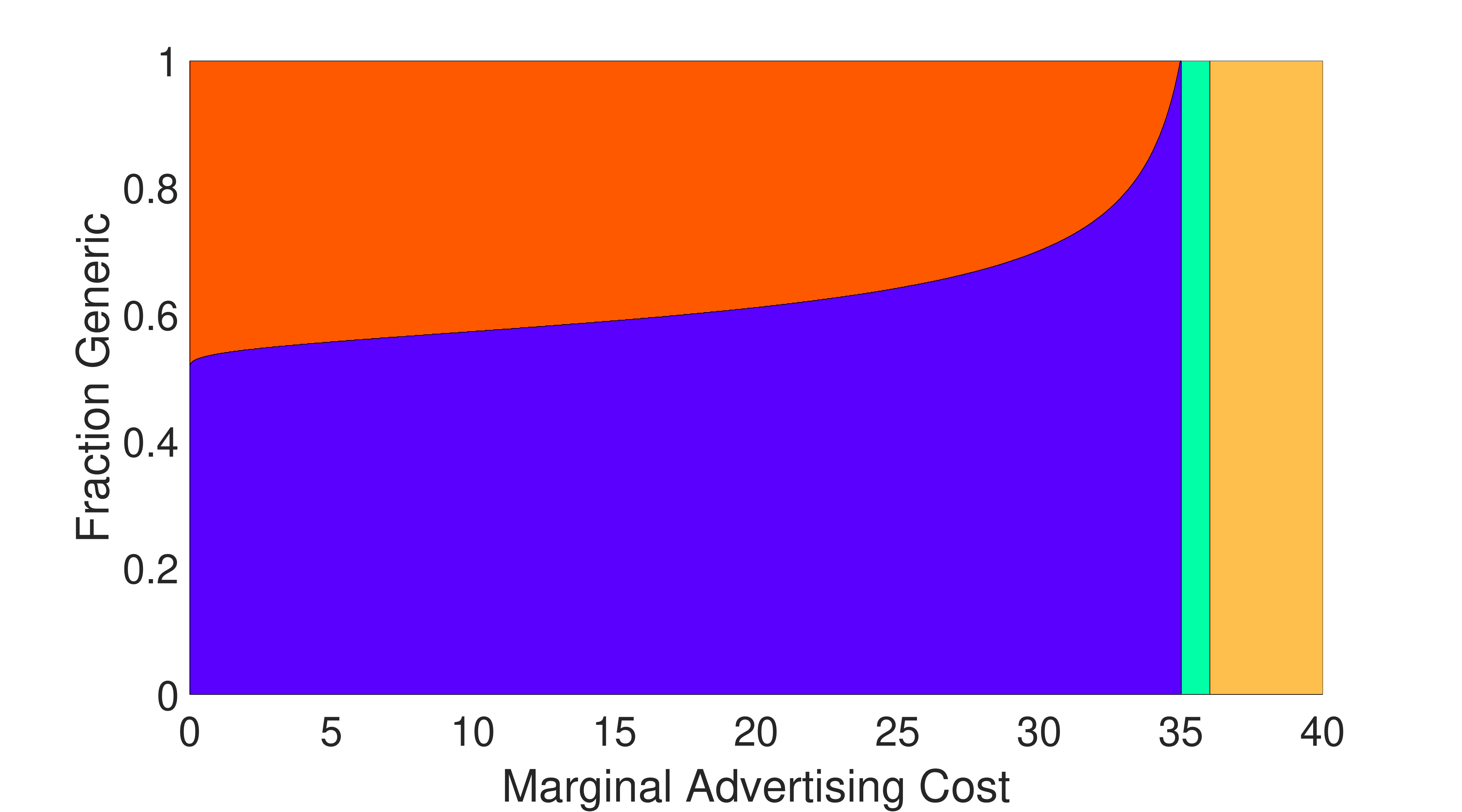}
    \caption{\textbf{Regions of existence and stability.} We numerically compute the regions of existence and stability outlined in Sections \ref{sec:exist_sm} and \ref{sec:stab_sm}. The blue area shows where the differentiated state exists and is stable, and the undifferentiated state fails to exist. The red area shows the region where neither the differentiated state nor the undifferentiated can exist; in this case, a bimodal initial condition will result in firms moving from the generic group to the name-brand group until a feasible fractionation is reached. The green/teal area shows a region of bistability, where both the differentiated state and the undifferentiated state exist and are stable (any fractionation is possible for the differentiated state within this region). The orange area shows where only the undifferentiated state exists. In all simulations, we set $N = 1000$, $ \kP = \kQ = \nu= \mu = \width = 1$, and $\Qmin = \kAD = 10$.} 
    \label{fig:genfracvsmargcost}
\end{figure}

We derived expressions for the self-consistent value of $a_{\textrm{name}}$ and the critical fractionation $\xcrit$ of generic firms analytically in the main text. For the sigmoidal case this is not feasible (it requires finding roots of a trancendental equation), but one can calculate these values numerically. We then are able to draw regions of existence of the differentiated state for a given fractionation $x$ and marginal advertising cost $\ka$: see Figure \ref{fig:genfracvsmargcost}.


\subsection{Existence of the differentiated state}\label{sec:exist_sm}

In Section \secModel, for the piecewise-linear advertising payoff function, we argued that the differentiated state may only exist when $\ka$ intersects $\MB(a_i \vert \vec{a})$. This holds in the current sigmoidal case as well, so the following is a necessary condition for the existence of a differentiated state:
\begin{equation}
    \ka < k^{crit}(\width,N,\Qmin,\kAD,\kQ,\kP) = \max_{a} \MB(a\vert\vec{a})\;.
    \label{eq:bimodNeces}
\end{equation}
Here $\max_{a} \MB(a\vert\vec{a})$ can be expressed algebraically as 
\begin{align}
    &\max_{a} \MB(a\vert\vec{a}) =\dfrac{4 (N-1)}{27 N \width \kAD \kP}\bigg(   \Qstar(\abar) \sqrt{\kAD^2+[\min \Qstar ]\kAD+[\min \Qstar]^2} + \frac{1}{2}\kAD^2 \nonumber \\
    &-[\min \Qstar ]\kAD -[\min \Qstar]^2 \bigg)\bigg( \Qstar(\abar) + \left.\frac{1}{2}\sqrt{\kAD^2+[\min \Qstar ]\kAD+[\min \Qstar ]^2} \right), \nonumber 
\end{align}
where $\min\Qstar  =\Qmin - \kQ \kP$, and $\Qstar(\abar) = \kAD/2 +\Qmin - \kQ \kP$. This condition appears to be sufficient as well as necessary (see Section \ref{sec:stab_undiff_state}).

\subsection{Stability of the differentiated state}\label{sec:stab_sm}

We investigate the stability of the differentiated state by examining the functional form of $da_i/dt$, which we can reduce to the following form:
\begin{equation}
    \frac{da_i}{dt} = -\alpha \tanh^3 \left( \dfrac{a_i-\abar}{\width} \right) - \beta \tanh^2 \left( \dfrac{a_i-\abar}{\width} \right) +
    \alpha  \tanh \left( \dfrac{a_i-\abar}{\width} \right) + \beta - \ka\;,
    \label{eq:tanhcubic}
\end{equation}
where $\alpha = (N-1) \kAD^2/(8\width N \kP)$ and $\beta = (N-1) \kAD \left( 0.5 \kAD -\kQ \kP \right)/(8 \lambda N \kP) $. 
Factoring out $\alpha$ from $\eqref{eq:tanhcubic}$ gives
\begin{equation}
    \frac{da_i}{dt} = \alpha\left(- u^3 -\gamma u^2 +  u +\gamma - k^*_a \right)\;, \label{eq:dadt_u}
\end{equation}
where $u = u(a_i\vert \vec{a}) = \tanh\left[(a_i-\abar\right)/\width]$, $\gamma = \beta/\alpha$, and $k^*_a = \ka/\alpha$. Because we are only interested in fixed points, we can ignore the $\alpha$ outside of the parenthesis as this just sets a time scale for equilibration. 

We define $g[u(a_i\vert \vec{a})] = - u^3 -\gamma u^2 +  u +\gamma - k^*_a$. To assess linear stability, we show that the eigenvalues of the Jacobian matrix of our dynamical system are negative in a similar fashion to the method used in reference \cite{clifton2016handicap}.  One can show that $\partial/\partial a_j (da_i/dt)$ scales like $1/N$ when $i\neq j$ and hence that these terms can be ignored when determining stability for $N \rightarrow \infty$.  Now, consider the main diagonal of the Jacobian of our system as $N \rightarrow \infty$:
\begin{equation}
    \frac{\partial}{\partial a_i} \left(\frac{da_i}{dt}\right) = g'[u(a_i\vert \vec{a})] \frac{\partial u}{\partial a_i}
    \label{eq:stabcond}
\end{equation}
where
\begin{equation}
    \frac{\partial u}{\partial a_i} =  \sech^2\left[\frac{a_i-\abar}{\width}\right] > 0\;.
\end{equation}
Hence, the sign of $\partial/\partial a_i (da_i/dt)$ is determined by the sign of $g'[u(a_i\vert \vec{a})]$. Since stability must alternate in one dimensional flows, the only stable nonzero fixed point must be the far right intersection between $\MB$ and $\ka$ (see \FigMid \advdynamics) located at $a=a_{\textrm{name}}$. We can write $g'[u(a_{name}\vert \vec{a})]$ as 
\begin{equation}
    g'[u(a_{name}\vert \vec{a})] = -\dfrac{1}{12} w^{2/3} -\dfrac{1}{3}v - \dfrac{4}{3} \dfrac{v^2}{w^{2/3}} \, ,
    \label{eq:stabcond2}
\end{equation}
where 
\begin{align}
    v &= \gamma^2 +3, \\
    w &= 72\gamma -108\ka^*-8\gamma^3 + 12 \sqrt{12 \ka^*\gamma^3 -12 \gamma^4 +81 (\ka^*)^2 -108 \ka^* \gamma +24 \gamma^2 -12}\;.
\end{align}
It can be shown that the expression in Eq.~\eqref{eq:stabcond2} has a maximum of zero which occurs at one point, specifically,
\begin{equation}
    \ka^*= \dfrac{2}{27} \gamma (9-\gamma^2) + (\gamma^2+3)^{\frac{3}{2}}\;,
    \label{eq:NeuStable}
\end{equation}
or, expressed in terms of the original parameters, 
\[ 
    \ka = \max_{a} \MB(a\vert\vec{a}).
\]
This also happens to be the threshold for existence of the differentiated (bimodal) state. So, as long as the bimodal state exists, the Jacobian has negative values along its main diagonal and, therefore, it is a stable solution as $N\rightarrow \infty$.

\subsection{Stability of the undifferentiated state} \label{sec:stab_undiff_state}

To examine stability of the undifferentiated state, we set $a_i = \abar = 0$, for all $i$. We again assume $N\rightarrow \infty$, and thus any change in $\abar$ is negligible under single firm perturbations. If there is an interval of positive perturbations neighboring zero for which $da_i/dt < 0$, then the undifferentiated state is stable under single firm perturbations for sufficiently small perturbations. If $a_i = \abar = 0$ then $u(0 \vert \vec{0}) = \tanh\left[(a_i-\abar\right)/\width]= \tanh(0) =0$. Hence, substituting $u = 0$ in to \EqMid \eqref{eq:dadt_u} gives
\begin{equation}
    \frac{da_i}{dt} = \alpha(- u^3 -\gamma u^2 +  u +\gamma - \ka^*) =  \alpha(\gamma -\ka^*) < 0\;.
\end{equation}
By continuity of $da_i/dt$, there must exist an interval $(0,\epsilon)$ such that $da_i/dt <0$. Since $\alpha >0$, the statement $k^*_a > \gamma$ is sufficient to ensure stability of the undifferentiated state. Taken together with Eq.~\eqref{eq:NeuStable} above, we observe that a region of bistability exists where both the undifferentiated state and the differentiated state are stable:
\begin{equation}
    \gamma < \ka^* < \dfrac{2}{27} \gamma (9-\gamma^2) + (\gamma^2+3)^{\frac{3}{2}}.
    \label{eq:BiStable}
\end{equation}
In the original parameters with $N \rightarrow \infty$, \EqMid \eqref{eq:BiStable} can be written as
\begin{align}
    \MB(\abar \vert\vec{a}) = \dfrac{  \kAD [\Qmin+\kAD/2-\kQ \kP] }{4 \width \kP} < \ka <\max_{a} \MB(a\vert\vec{a})\;.
    \label{eq:BiStable_Norm_sig}
\end{align}
In more intuitive terms this can be expressed as
\begin{equation}
\left.\dfrac{\partial \pi_i}{\partial a_i}\right\vert_{a_i =\abar}  < 0 < \max_a \dfrac{\partial \pi_i}{\partial a_i}\;.
\end{equation}
This result is identical to the one given in \EqMid \eqInuitStab.

\begin{figure}[t!]
    \centering
    \includegraphics[width=110mm]{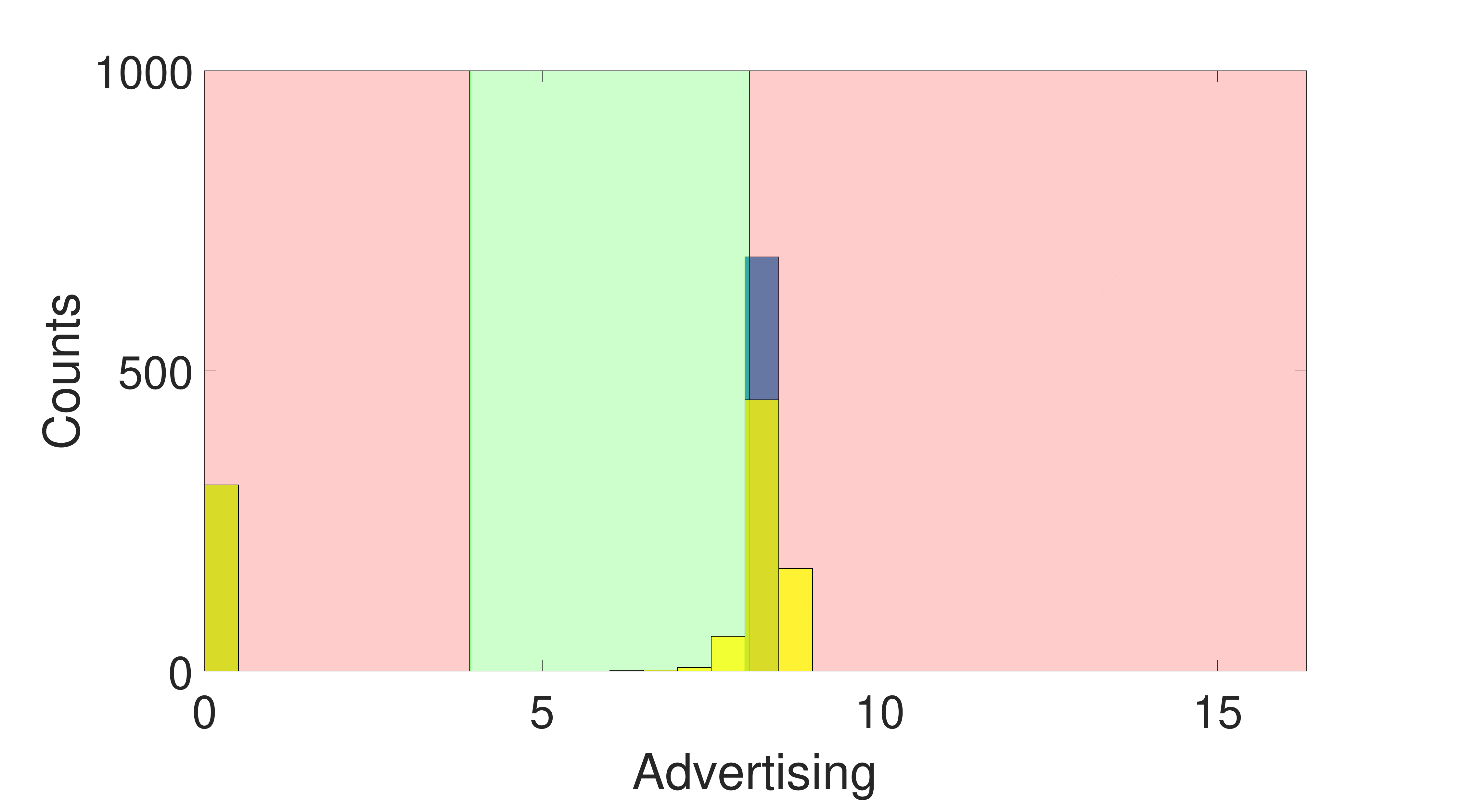}
    \caption{\textbf{Simulation with heterogeneous firms.} In this figure we compare the steady state of the system with heterogeneous firms (yellow, for firms with nonuniform $\kAD$) to the steady state with homogeneous firms (blue). The equilibrium with heterogeneous firms is qualitatively similar to the equilibrium with homogeneous firms, but the nonzero-advertising peak representing name-brand firms is spread out. The background colors represent regions where advertising increases (green) and where advertising decreases (red) in the homogeneous case and are added to show that the system is in equilibrium. As in the main text, we set the number of companies to $N=1000$, $\ka = \kP = \kQ = \nu= \mu = \width = 1$, and $\Qmin = \kAD = 10$. For the heterogeneous firms, we added a random variable $\xi$ drawn from $\mathcal{N}(0,9)$ to $\kAD$. }
    \label{fig:Hetero_Advert}
\end{figure}

\section{Numerical exploration}

Because of the high dimensional parameter space, we do not fully explore all possibilities, but we focus on cases that seem to be most relevant to the real world.  In particular, we look at (1) production cost curves that offer economies of scale, where $\mu < 1$; (2) advertising cost curves that offer economies of scale, where $\nu < 1$; and (3) heterogeneous (rather than identical) firms with varying intrinsic demand ($\Qmin$) or varying demand response to advertising  ($\kAD$).

When $\mu$ and/or $\nu$ are less than one, the quantitative results (in particular the existence and price of the ``name-brand'' product) differ slightly from the case where $\mu = \nu = 1$, but are still in good qualitative agreement.

When we introduce heterogeneity among firms, we find that the bimodal equilibria show some nonzero spread of advertising for the name-brand firms, as opposed to the delta-function distribution seen in the case of identical firms.  The advertising budgets for name-brand firms are, however, still centered about the position of the delta spike predicted by our theory for identical firms---see Fig.~\ref{fig:Hetero_Advert}.

\section{Welfare analysis}

The existence and persistence of name brand goods is of intrinsic interest, but it is also of interest to determine if advertising is beneficial to society at large, and if advertising generates more profit for the industry as a whole. We address these questions by considering economic welfare, similarly to the approach taken in reference \cite{tremblay2012advertising}.  We define total welfare to be the sum of consumer welfare---the difference between the maximum price an individual is willing to pay for a good or service and the price they actually pay---and the total profit generated by the market. We use post-advertising preferences when calculating consumer welfare, effectively interpreting consumer demand \textit{after} advertising to reveal the true preferences. Further consideration of this point is merited to better understand how advertising should factor into consumer welfare, but we leave that for future work.

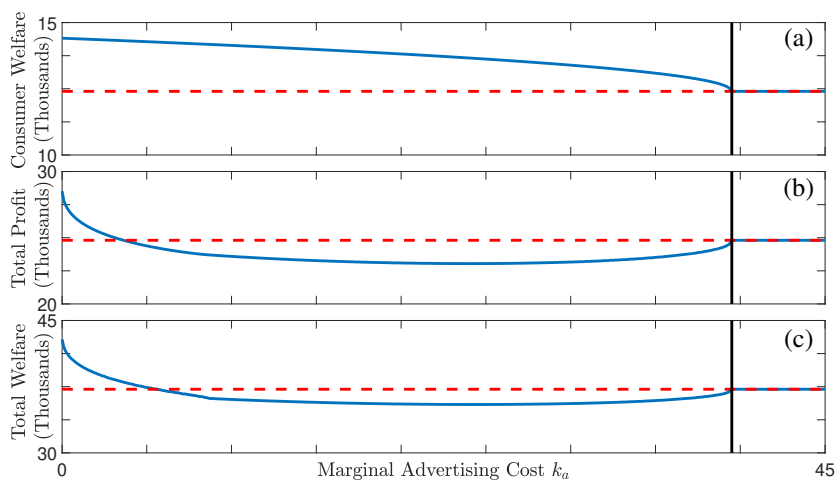
\begin{figure}[t!]
    \centering
    \begin{tikzpicture}
        \node at (0,0)    {\includegraphics[width=110mm]{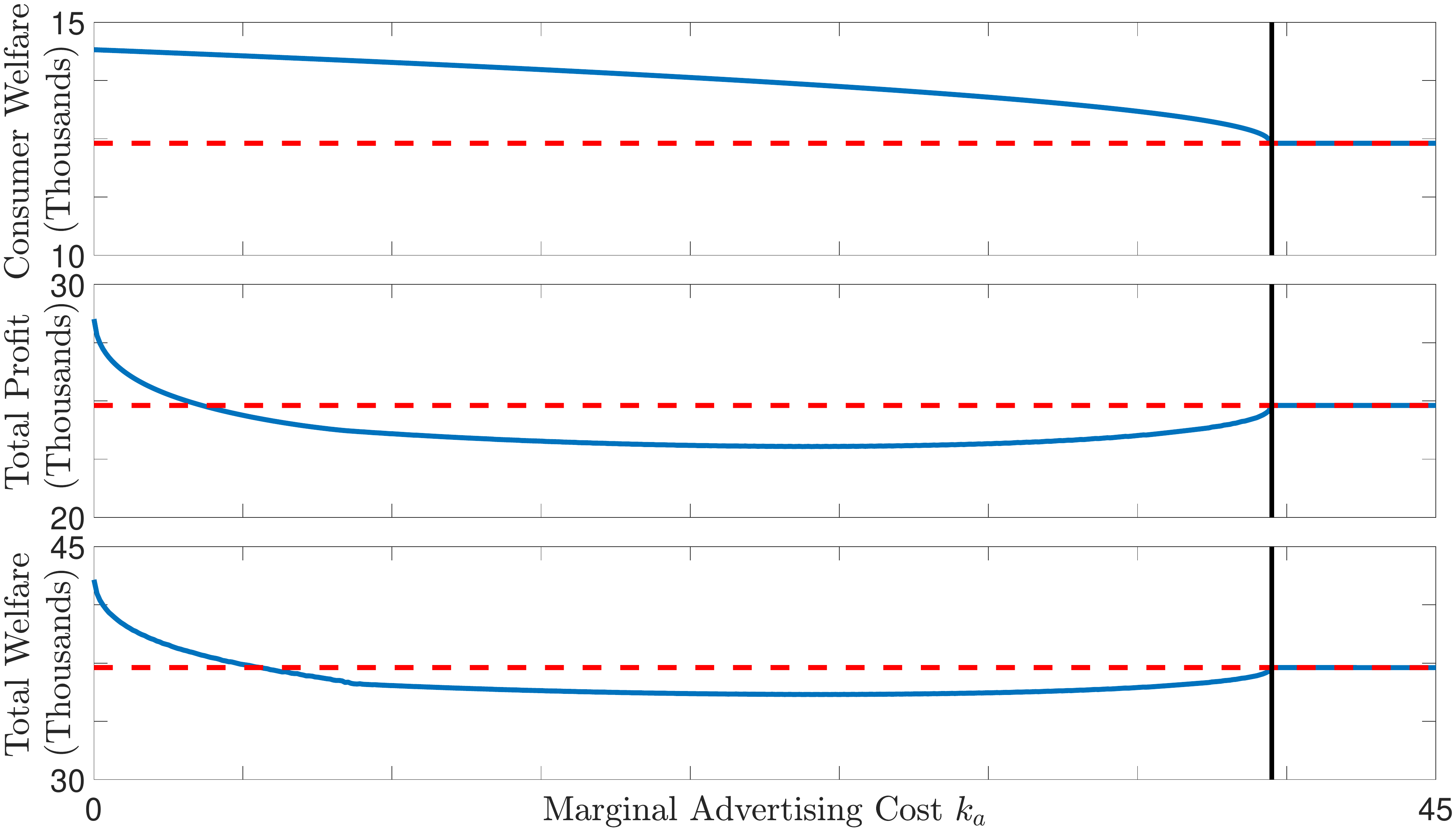}};
        \node at (5,2.75) {(a)};
        \node at (5,0.75) {(b)};
        \node at (5,-1.25) {(c)};
    \end{tikzpicture}
    \caption{\textbf{Optimized consumer welfare, profit and total welfare given marginal advertising costs.} In (a) we display the optimal total consumer welfare generated by the market across all possible differentiated equilibria for a given marginal advertising cost $\ka$ (blue, solid) and the total consumer welfare generated by the undifferentiated state (red, dashed). The black line in all three panels indicates when $\ka = \max \MB (a_i \vert \vec{a})$ and thus, past that point the differentiated state ceases to be stable. In (b) we display the optimal total profit generated by the market across all possible differentiated equilibria for a given marginal advertising cost $\ka$ (blue) and the total profit generated by the undifferentiated state (red, dashed). In (c) we display the optimal total welfare generated by the market across all possible differentiated equilibria for a given marginal advertising cost $\ka$ (blue, solid) and the total welfare generated by the undifferentiated state (red, dashed). We set $N= 100$, $\Qmin =30$, $\kAD =5$,  and $\width =\mu =\nu =\kQ= \kP= 1$, with $\ka$ ranging over a range 0 to 45.}
    \label{fig:frac_advert}
\end{figure}

\FigBeg \ref{fig:frac_advert}(a) displays the maximum possible consumer welfare (optimizing over all fractions generic in differentiated equilibria) for a given marginal advertising cost $\ka$ (blue, solid) and the total welfare for the undifferentiated state (red, dashed). The black line indicates when $\ka = \max \MB (a_i \vert \vec{a})$---past that point the differentiated state ceases to be stable. The maximum possible consumer welfare for the differentiated state is greater than the undifferentiated state across all $\ka$.

\FigBeg \ref{fig:frac_advert}(b) shows the maximum possible total profit in the market (optimized over all stable fractions generic in differentiated equilibria) for a given marginal advertising cost $\ka$ (blue, solid) and the total profit for the undifferentiated state (red, dashed). The differentiated state is more lucrative (at the industry-wide level) when marginal advertising costs are low, and the undifferentiated case is more lucrative for high marginal advertising costs.

\FigBeg \ref{fig:frac_advert}(c) shows the maximum possible total welfare in the market. Note this is not simply the sum of the numbers in panels (a) and (b), as they may correspond to different stable fractionations $x$. The differentiated state (blue, solid) is more beneficial when considering the interests of consumers and producers when marginal advertising costs are low, and the undifferentiated case (red, dashed) is more beneficial when marginal advertising costs are high. In this numerical experiment we set $N= 100$, $\Qmin =30$, $\kAD =5$,  and $\width =\mu =\nu =\kQ= \kP= 1$, with $\ka$ ranging over a range 0 to 45.

According to Figure \ref{fig:frac_advert}, the differentiated state produces more total welfare than the undifferentiated state when the marginal cost of advertising is low. Conversely, the undifferentiated state generates greater total welfare when the marginal cost is relatively high. We leave analytical exploration of the effect of advertising on welfare and profit as a possible extension to this work.

\section{Comparing model predictions and data} \label{sec:compare_model_data}

\begin{figure}[ht]
    \centering
    \begin{overpic}[width=88mm]{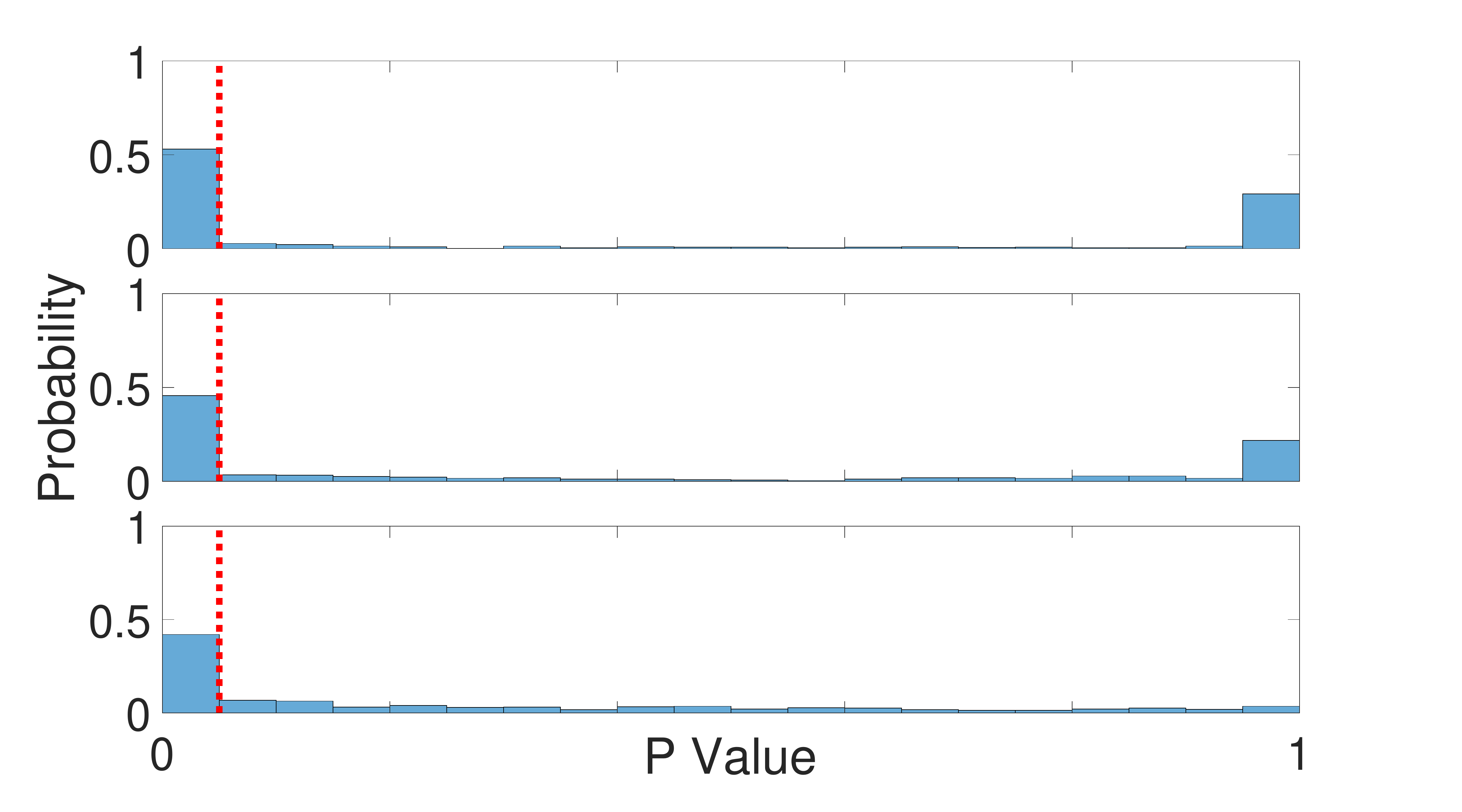}
        \put(83,48)  {(a)}
        \put(83,32)  {(b)}
        \put(83,16)  {(c)}
    \end{overpic}
    \caption{\textbf{Histograms of Fitting Statisitics.} Distributions of $p$ values from tests for unimodality (a-b) and consistency of price data with model predictions (c). The red dashed line indicates the significance level of $.05$. Top and middle row:  LUU and Hartigan's Dip Test (respectively).  Rejection (low $p$-value) means the price distribution is not consistent with unimodal null hypothesis. Bottom row: KS test. Rejection (low $p$-value) means the price distribution is not consistent with same-distribution null hypothesis (i.e., model and data not from same distribution).}
    \label{fig:Hart_Dip_LUU}
\end{figure}

In Section \secStatTests, we describe statistical tests used to assess how our model's predictions compare to real-world data. We use the Kolmogorov-Smirnov test (KS test) to determine if the sample data and the model predictions are likely to have come from same distribution. We use Hartigan's Dip Test to see if we could reject unimodality of the sample data. We display the resulting distributions from the application of these statistical tests in \FigMid \ref{fig:Hart_Dip_LUU}.

Hartigan's Dip Test appears to be ``excessively conservative and insensitive'' at small sample sizes according to \cite{clifton2016handicap}. Clifton et al.~used an altered test, the Least Unimodal Unimodal (LUU), to develop a more sensitive bootstrapped dip statistic to test for unimodality.

According to the LUU Test $53\%$  of price distributions are inconsistent with unimodality at a significance level $0.05$ compared to $46\%$ of price distributions  inconsistent with unimodality when applying Hartigan's Dip Test. For $58\%$ of our model fits to the top 500 products, we fail to reject the null hypothesis at a significance level of $0.05$: the data and the model prediction may come from the same distribution.

\section{Seeding the minimization algorithm} \label{sec:seed_alg}

To seed the initial choice of parameters for fitting, we attempt to match the modes of the model distribution, $f(p)$, to the modes of the empirical distribution $g(p)$ inferred from the price data. Since our model only produces unimodal or bimodal distributions, we produce kernel density estimates (KDEs) of the data with increasing bandwidths until we detect at most two peaks at prices $P_1$ and $P_2$, where  $P_2 > P_1$. Numerically, we accomplish this using KDE bandwidths increasing progressively from $0.1$ to $10$ in increments of $0.1$. We apply a threshold to peak-detection such that only peaks of height greater than $0.1$ are counted. 

If the KDE is bimodal then we wish to choose seed values such that the two modes of the differentiated state (the generic brand price and name brand price) match $P_1$ and $P_2$. We cannot easily predict the equilibrium advertising value for name brands $a_{\textrm{name}}$, so $\Qfree$ (and hence $\Pstar$) cannot be set exactly to match $P_1$ and $P_2$. Instead, we make the assumption that the average advertising level $\abar$ is far away enough from $0$ and $a_{\textrm{name}}$ that $\Qfree$ is near its saturation values at $\pm \infty$ for generics and name brands---meaning that name brands receive the maximum demand increase due to advertising and generics receive no increase in demand due to advertising.

We minimize the magnitude of profit's derivative with respect to price when $\Qfree = \kAD +\Qmin$ and $\Qfree = \Qmin$. This leads to the objective function 
\begin{equation}
     R(\kAD, \kP,\kQ,\Qmin,\mu)=\left\vert \dfrac{\partial \pi_i}{\partial P_i} \right\vert_{P_i = P_1,a_i\rightarrow-\infty}+  \left\vert \dfrac{\partial \pi_i}{\partial P_i} \right\vert_{P_i = P_2, a_i\rightarrow \infty}.\label{eq:opt}
\end{equation}
$R(\kAD, \kP,\kQ,\Qmin, \mu)$ is minimized when $\partial \pi_i/\partial P_i\vert_{P_i = P_1,a_i\rightarrow-\infty}$ and $\partial \pi_i/\partial P_i\vert_{P_i = P_2,a_i\rightarrow\infty}$ are both zero. When $R(\kAD, \kP,\kQ,\Qmin,\mu)$ is minimized the optimal price for name brands is $P = P_2$ (in the limit $a \rightarrow +\infty$) and the optimal price for generics is $P = P_1$ (in the limit $a \rightarrow -\infty$). We minimize $R$ using the Nelder-Mead algorithm \cite{nelder1965simplex} and the resulting values for $\kAD, \kP,\kQ,\Qmin,$ and $\mu$ are used as seeds for fitting our model to the data. We set $\kAD = 1.1 \times 10^7, \kP = 4.4 \times 10^6,\kQ = 100, \Qmin = 2.5 \times 10^7,$ and $\mu = 0.4$ to seed the $R$-minimization problem (where these values are derived from a more complete exploration of the parameter space in the case of a single product from the data set). 

The initial proportion of generic firms $x$ is set by measuring the percentage of price entries that are less than the quantity  $P_1 + w$, where $w$ is the peak width at half-prominence for the peak at $P_1$. The initial advertising distribution is given by $x\delta(a_i -1) + (1-x)\delta(a_i-100)$, where $\delta$ represents the Dirac delta function.

If the KDE is unimodal (i.e., $P_1 = P_2$), the we take a different approach.  There is a critical cost of advertising $\ka^{\textrm{crit}}$ defined by $\ka^{\textrm{crit}} = \nu^{-1} (a_{\max})^{1-\nu} \MB(a_{\max}\vert \vec{a})$, where for $\ka > \ka^{\textrm{crit}}$, $da_i/dt < 0$ for all $i$ and no bimodal state exists, since there will be no intersection between $B$ and the advertising cost curve $C'_a(a) = \nu \ka a^{\nu-1}$ (we assume $0 < \nu \leq 1$).  Here $a_{\max}$ is defined as the advertising value such that $\MB(a_{\max}\vert \vec{a}) = \max_a \MB(a_i\vert \vec{a})$, i.e., the arg max of the function $B$.

Note that this critical value $\ka^{\textrm{crit}}$ is state-dependent.  We set the initial advertising distribution for this undifferentiated case to $\delta(a_i-100)$, so $a_i=\abar=100$ for $i=1,\ldots,N$.  Given this choice, we found that setting $\ka =  \nu^{-1} (a^*)^{1-\nu} \MB(a_{\max} \vert \vec{a})$ was sufficient to ensure that the system evolved to a unimodal price distribution, where $a^*$ is any sufficiently large value above $a_{\max}$ (the threshold for ``sufficiently large'' can be calculated from model parameters).  We used $a^* = 300$ in our simulations.

Additionally, for the undifferentiated state, we redefine $R$ as
\begin{equation}
     R(\kAD, \kP,\kQ,\Qmin,\mu)=\left\vert \dfrac{\partial \pi_i}{\partial P_i} \right\vert_{P_i = P_1,a_i =\abar=0}.\label{eq:opt_uni}
\end{equation}
We then minimize $R$ to set the parameters so the price in the undifferentiated state matches $P_1$.  We implement the Nelder-Mead algorithm with the same seeds as in the previously described differentiated case. 

There are three parameters we don't explicitly fit, namely, $\ka$, $\lambda$ and $\nu$  (two in the undifferentiated case, since $\ka$ is specified). For these we use the same values employed to seed $R$; the values result from a more thorough exploration of the parameter space in the case of a single product from the data set.  They are: $\ka \approx 9100$, $\lambda \approx 9.7$ and $\nu \approx 0.57$.  We chose not to fit these parameters for every product because price distributions appeared to be insensitive to their values.  Given the inherent limitations of our ``toy'' model, our goal was to demonstrate that agreement between model and data is plausible, not to exhaustively discover the best possible agreement.

\section{Data adjustment} \label{sec:data_adjustment}

On examining the data, it becomes apparent that there is bias present toward prices ending in certain digits. Prior research has highlighted this phenomenon in pricing of goods, known as \textit{psychological pricing} \cite{jovsic2018application,el2005price,schindler2009patterns,stiving1997empirical}. Since our model does not take into account this type of bias, we develop a method to remove this bias from the price distribution.

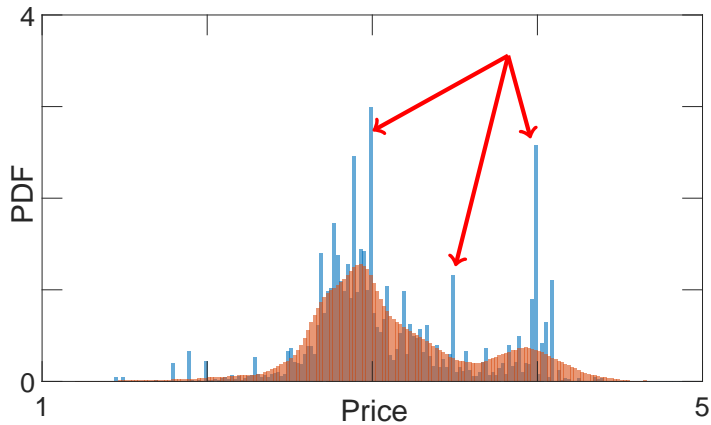
\begin{figure}[th]
    \centering
    \begin{tikzpicture}
    \node at (0,0) {  
        \includegraphics[width=110mm]{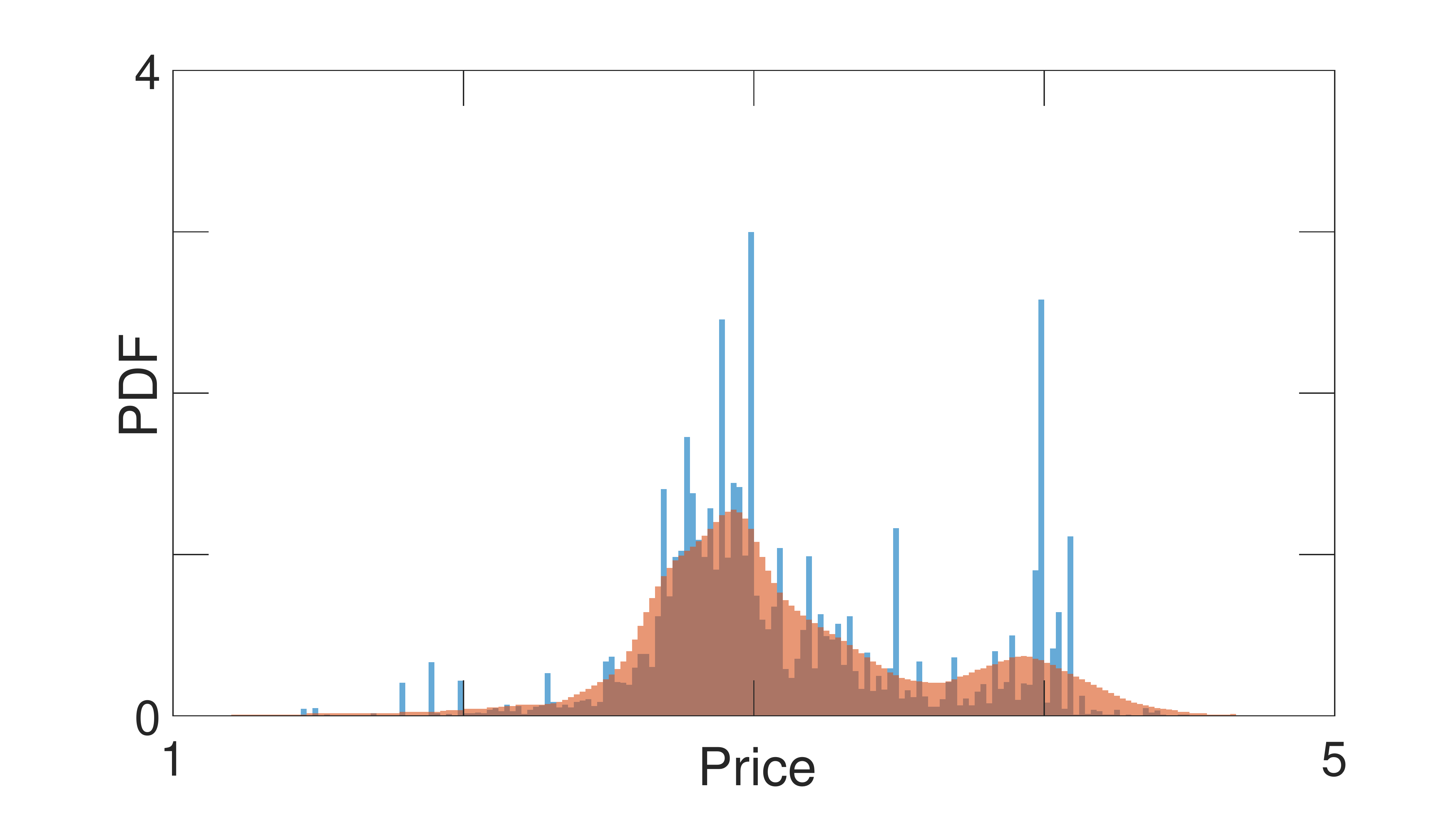}};
    \draw [->,red,ultra thick] (2,2) -- (0.2,1);
    \draw [->,red,ultra thick] (2,2) -- (2.3,0.9);
    \draw [->,red,ultra thick] (2,2) -- (1.3,-0.8);
    \end{tikzpicture}
    \caption{\textbf{Raw and Smoothed Price Data.} Example of the raw price data with spikes situated near certain ending digits (blue) and the smoothed, debiased data that retains the quantitative properties of the underlying distribution while removing the spikes (orange). Red arrows point to example spikes located at $\$2.99$, $\$3.49$, and $\$3.99$.}
    \label{fig:RawDataMilk}
\end{figure}

This bias towards certain ending digits creates spikes in each product's price distributions, as displayed in \FigMid \ref{fig:RawDataMilk} (blue). One might consider using a kernel density estimate (KDE) \cite{parzen1962estimation} or a filter to remove these spikes \cite{mohammad2012filtering}. However, with a KDE approach, the amount of smoothing required to effectively remove these spikes significantly alters the underlying distribution. With the filtering approach, the ``probability mass'' of the spike is redistributed nonlocally across the entire distribution, again altering it in an undesirable way. We develop a new method that attempts to remove spikes while keeping the underlying distribution largely unchanged. An example applying our method to data can be seen in \FigMid \ref{fig:RawDataMilk} (orange).
 
Details of our ``debiasing'' method are reported in an upcoming paper \cite{johnson2019bias}, however we also explain the idea briefly here.  We begin by defining a metric, $B_1(p)$, for how biased a distribution is towards a price value $p$. We calculate $B_1(p)$ as follows:
\begin{equation}
    B_1(p) =\dfrac{\rho_0(p)}{\rho_{1}(p)},
\end{equation}
where $\rho_0(p)$ is the value of the normalized histogram at $p$ and $\rho_1(p)$ is a kernel density estimate of $\rho_0$ using a Gaussian kernel with a narrow bandwidth (e.g., here we use $h=0.01$).  Conceptually similar to generation of a KDE, we then replace the data point at each price $p_i$ with a Gaussian that has variance $s B_1(p_i)^{r}$. This generates a new price distribution $\rho_2(p)$, where the parameters $s>0$ and $r>0$ determine the natural amount of smoothing when there is no bias and how the smoothing scales with bias---linearly, sublinearly, or superlinearly---respectively.

When we apply this method, we set $r = 0.5$ and set $s = \sigma (4/3 n)^{-0.2}$ based on Silverman's ``rule of thumb'' \cite{silverman1986density}, with $n$ being the number of data entries and $\sigma$ defined as the standard deviation of the data.   We iterate this process with $B_i(p)$ being defined as
\begin{equation}
    B_i(p) =\dfrac{\rho_0(p)}{\rho_{i}(p)}
\end{equation}
until we arrive at a distribution $\rho_{\infty}$, such that it is fixed under this iteration. 

As the sample size goes to infinity, the method is equivalent to the following integral transform:
\begin{equation}
     \rho_{i+1}(p) = \int_{-\infty}^{\infty} \rho_0(p') \   \mathcal{N}(p',s (B_{i}(p-p'))^{r}) dp'.
\end{equation}
Letting $i \rightarrow\infty$ yields the smoothed distribution 
\begin{equation}
     \rho_{\infty}(p) = \int_{-\infty}^{\infty} \rho_0(p') \   \mathcal{N}(p',s (B_{\infty}(p-p'))^{r}) dp'. 
\end{equation}
The distribution $\rho_{\infty}$ is fixed under this integral transform. As mentioned above, tests of how effectively this method performs ``bias'' removal are reported in an upcoming manuscript \cite{johnson2019bias}. We move forward assuming that this method works sufficiently well; this assumption is supported by qualitative visual examination of distributions before and after this ``debiasing'' procedure\footnote{Note that the KDE performs a uniformly diffusive smoothing operation, whereas our approach assumes that psychological bias in pricing is better corrected for by a locally varying redistribution of probability mass}.

\section{Data Availability}

The data that support the findings of this study are available from The Nielsen Company (US), LLC but restrictions apply to the availability of these data, which were used under license for the current study, and so are not publicly available. Data are however generally available for scientific research with an institutional or individual subscription \cite{Kilts2014Nielsen}.

\begin{footnotesize}
	\textit{Researcher(s) own analyses calculated (or derived) based in part on data from The Nielsen Company (US), LLC and marketing databases provided through the Nielsen Datasets at the Kilts Center for Marketing Data Center at The University of Chicago Booth School of Business.}
\end{footnotesize}

\begin{footnotesize}
	\textit{The conclusions drawn from the Nielsen data are those of the researcher(s) and do not reflect the views of Nielsen. Nielsen is not responsible for, had no role in, and was not involved in analyzing and preparing the results reported herein.}
\end{footnotesize}

\bibliographystyle{siamplain}
\bibliography{advert}